# 100 Years of Quanta:
## Complex-Dynamical Origin of Planck's Constant and Causally Complete Extension of Quantum Mechanics

A.P. KIRILYUK*

Solid State Theory Department, Institute of Metal Physics
36 Vernadsky Avenue, 03142 Kiev-142, Ukraine

ABSTRACT. On 14 December 1900 Max Planck first formulated the idea of intrinsic discreteness of energy of solid-body oscillators and expressed the discrete energy portions, or quanta, as the product of frequency of emitted or absorbed radiation and a new universal constant now known as Planck's constant. Despite the following spectacular progress of thus initiated "quantum mechanics" (and "new physics" in general), the physical origin of both energy discreteness and universality of Planck's constant, determining quantization of very diverse object behaviour, remain mysterious, as well as other "peculiar" properties of quantum dynamics. In this paper we review the recently proposed, causally complete extension of quantum mechanics consistently explaining all its "mysteries", including action and energy quantization, by the irreducibly complex, "dynamically multivalued" behaviour of the underlying simple, physically real system of two interacting protofields (quant-ph/9902015, quant-ph/9902016). We emphasize the truly fundamental and realistic character of the theory containing no imposed "postulates", "principles", or inserted "entities" except one, unavoidable (and mild) assumption about the qualitative, physical nature of the protofields. All the observed entities and their properties, starting from physically real space, time, and elementary particle structure, are consistently derived, in exact correspondence with their emergence in real, irreducibly complex system dynamics (physics/9806002). The latter provides also natural (dynamic) unification of the causally extended versions of quantum mechanics, relativity, and field theory, including unification and causal understanding of particle interaction forces. Intrinsic realism and completeness of the obtained world picture are in agreement with the "absolute reality" quest of Max Planck and actually confirm his famous doubts about the conventional, abstract and formally postulated scheme of quantum mechanics (cf. quant-ph/9911107, quant-ph/0101129). We outline various applications of the obtained results providing many independent confirmations of the theory and successful solutions to numerous fundamental and practical problems dangerously stagnating within the canonical, dynamically single-valued approach that continues to dominate in science because of purely subjective influences emphasised in the "scientific revolution" description by Thomas Kuhn.

---

*Address for correspondence: A.P. Kirilyuk, Post Box 115, 01030 Kiev-30, Ukraine. E-mail: kiril@metfiz.freenet.kiev.ua.

CONTENTS

A.P. KirilyukA.P. Kirilyuk

## 1. Introduction: Max Planck's absolute reality and the new physics

One hundred years have passed now since the conventional "birth" of quantum mechanics on the 14th December of 1900 when Max Planck, the 42-year-old professor of the University of Berlin, presented a report at a meeting of German Physical Society where he specified the idea of fundamental discreteness of energy emitted and absorbed, in the form of electromagnetic waves, by any individual microscopic oscillator, or "resonator", within a "black body", which is an isolated system of solid body and its radiation maintained at certain temperature by the stationary (equilibrium) energy exchange between the body's oscillators and radiation [1]. This first explicit, well specified appearance of natural discreteness (or quantization) of energy did not create any remarkable resonance in the scientific community at that time and actually was a "frustrating" assumption of Max Planck difficult for him (see e. g. [2-4]), since he was "forced" to make it as the only possible "physical" explanation for the formula for energy spectrum of the black body radiation that was more formally obtained (partially "guessed") by him shortly before that [5] by comparison of thermodynamical analysis results with both Wien's law (derived in 1886) and experimentally observed deviations from it at lower frequencies (that agree with the Rayleigh-Jeans law independently derived by W. Rayleigh in the same year 1900). The first cry of the whole "new physics" [6,7] was thus hardly heard by the scientific community, even though its further growth within the first three decades of the 20th century, including the correlated "explosive" emergence of quantum mechanics, special and general relativity, field theory, and cosmology, remains one of the most intensive and spectacular knowledge revolutions.

However, the accomplishments of a hundred years of science development separating us from Planck's "undesired" child should not be exaggerated either, and further evolution of energy quanta hypothesis provides itself the best example of the intrinsic weakness of purely abstract, mechanistic way of the "new" science development imposed by certain its later, somewhat too "prodigious" promoters, often against the desire of original founders of new ideas, including Max Planck, who considered that any true scientific progress can only increase realism and consistency (or causal completeness) of knowledge [2,6,8]. Indeed, the basic assumption about the intrinsic discreteness of radiating oscillator (and any microscopic system) dynamics that gave rise to major doubts of its creator by contradiction to the "default" continuity of the "classical" world picture remains, within the conventional science, as poorly justified and "odd" today as it was at the moment of emergence one hundred years ago. The detailed scheme of quantum mechanics elaborated later *only postulates*, in various ways, but *without* any causal explanation, its key, properly "new" assertions, including the quantized character of observed quantities, which is *universally* determined by the fundamental action unit, Planck's constant $h$, introduced, together with its empirically specified value, in the original Planck communication [1]. The basic idea of microscopic oscillator energy discreteness is accompanied in [1] by the equally revolutionary assumption that each of the quantized portions of energy, $\varepsilon$, of emitted and absorbed electromagnetic radiation is proportional to its frequency, $\nu$, with the constant $h$ being the *universal* coefficient that relates the two quantities, $\varepsilon = h\nu$. Both this relation and the ensuing idea of photon as the physically specified quantum of electromagnetic radiation remain causally unexplained and even more contradictory than other "quantum mysteries" concerning massive particle behaviour. In this sense, it is difficult not to acknowledge today that Max Planck was a "*reluctant* revolutionary" [4] for the *right* reason, and the modern huge amplification of debates about the basis of the "new physics" and increasingly interested reconsideration of the century-old "puzzles", in direct connection to practical science problems [9-17], only confirm the major incompleteness of canonical science conventions that had provoked serious, and fully justified, Planck's doubts from the very beginning of their apparently "successful" establishment.

Discreteness of mechanical action, always changing by portions of $h = 6.6262 \times 10^{-27}$ erg·s, for the whole variety of elementary particles, compound (including macroscopic) quantum systems, and their properties (as diverse as nuclear, atomic and condensed matter phenomena, or else spin-related properties), is the central point of multiple manifestations of quantum-mechanical discreteness of any observable quantities, like energy and momentum, and therefore should also be directly related to the discrete structure of elementary material "bricks" of the world (elementary particle spectrum) and their intrinsic properties (such as rest mass, electric charge, and spin). This latter involvement of Planck's constant is specified, to a certain degree, within the idea of so-called "Planckian units" (of length, time, and mass-energy) playing a major role in particle physics constructions and first proposed, without any coincidence, by the same, "reluctant" (but *honest* and therefore *true*) revolutionary [18]. Neither the fundamental action discreteness, nor its astonishing universality, nor other related aspects of "quantum strangeness" (like "nonlocality", "duality", probabilistic "unpredictability", measurement "uncertainty", etc.) have ever been causally, physically explained by the conventional science, including its latest versions of the "theory of everything", despite innumerable pseudo-philosophical speculations and purely abstract, always formally imposed constructions, postulates and "principles". That major ambiguity in the very basis of conventional science world picture does not want to silently disappear behind visible successes of empirical applications of the postulated mathematical description, as many active proponents of the canonical quantum mystification seemed to hope (cf. "Copenhagen" and other "interpretations" in the scholar quantum mechanics), but on the contrary increasingly re-emerges today, around the next century border, as it can easily be seen from the current growing flux of works desperately tackling the same, "irresolvable" problems and the more and more evident impasse of the fundamental physics [19,20]. The resulting difficulties inevitably "propagate" to higher levels of the scholar picture of *de facto* unified real world, since even apart from the direct relation between neighbouring levels of world dynamics, the major deficiency of the *most fundamental*, quantum level certainly means that the *whole* conventional science misses "something essential" in its approach, which is simply more directly and "exactly" visible at the most elementary levels of dynamics.

Commemorating the 100th anniversary of quanta, and thus of the whole "new" physics, it is important to emphasize, rather than to hide, those problems in its modern state, as well as the fact that its pioneer, Max Planck, together with other true founders, Louis de Broglie [21] and Erwin Schrödinger [22], was strictly opposed to any anti-realistic, formal postulation of purely abstract "principles", or "laws of nature" and dangerous concessions to mysticism and inconsistent abstractions, which unfortunately dominated during the whole 20th century development of the fundamental science due to the well directed efforts of intrinsic adherents of the "mathematical physics" kind of imitation. The underlying difference in "moral principles" around the "acceptable" way of knowledge creation is also well illustrated by the firm logical and spiritual convictions of Max Planck [2,6,8] as compared to "fuzzy" values behind today's "post-modern" speculations of the "ironic science" [20] and shows quite clearly that any road of deviation from the unreduced truth/consistency and realism/causality leads inevitably to severe practical consequences for both science and its technological applications.

One cannot (and should not) stop the purely empirical development of technology, but without being seriously supported "from below" by the unreduced, causal understanding of reality, the technically powerful, but actually blind technology will inevitably touch directly the core of the unknown reality, with the real risk of equally deep, "global" kind of catastrophic destruction, and that is exactly the present-day situation in science/technology resulting from the superficial, "easy" attitudes to progress within the elapsing century of decadence. One certainly may be missing complete understanding of new data and should continue to look for it, but one must not replace it by a seemingly useful, "practically sufficient", but obviously incomplete imitation. That the creator of the hypothesis of quanta totally adhered to such attitude is clearly demonstrated by his "strangely" persistent doubts coming from their unexplained origin and contradiction to classical electrodynamics [2-4], even despite their quite successful appearance in his own work and convincing "experimental confirmation". The same unreduced causality requirement underlies the related rejection by Max Planck of reality of light quanta (photons) introduced by Albert Einstein in 1905 to account for effects of light interaction with matter in a situation essentially similar to the black-body radiation system, where neither the necessity of the radiation field discreteness, nor its detailed structure and origin were implied by the occurring processes (indeed, the physical nature of photons remains completely mysterious in the scholar science picture until now).[*] There are many other manifestations of unreduced realism and consistency in Max Planck's work and the way of doing science he defended, often with a reference to objectively existing, *absolute reality* independent of researchers and needing their ever growing understanding (see [2,6,8]).

## 2. Unreduced interaction complexity as the causally complete solution of quantum mysteries

### 2.1. Universal dynamic complexity and its relation to quantum behaviour

While the deep conflict between the unreduced reality and its abstract modelling in conventional fundamental science continues to grow ever since its explicit emergence in the "new physics" a century ago and now takes the form of a definite impasse of knowledge, or the "end of science" [20], accompanied by the blind technology domination, a qualitatively different, causally complete and well-specified solution to "unsolvable" problems of fundamental physics was recently proposed in the form of the new, *reality-based* concept of *dynamic complexity*, or "universal science of complexity" [23], that should be clearly distinguished from various abstract, non-universal and ambiguous *imitations* of complexity (see e. g. [24]) within the *same*, conventional science paradigm. The new, unreduced concept of dynamic complexity is based on the phenomenon of *dynamic redundance, or multivaluedness*, first discovered within theoretical description of a particular physical system of charged particle interacting with crystal and showing chaotic behaviour [25]. The results had much more general meaning and were then extended to progressively wider classes of systems incorporating general quantum chaos [26], quantum measurement and reduction for slightly dissipative quantum systems [27], and arbitrary real system of interacting entities with applications to particular cases from various levels of world dynamics, starting from elementary entities [21-23,28-30]. *Universality* of the dynamic redundance phenomenon, as well as accompanying *dynamic entanglement-disentanglement* mechanism of interaction development, the related concept of *dynamic complexity* and their mathematical description allows for application of the results obtained for quantum chaos and measurement cases to arbitrary system dynamics, which demonstrates the new level of unification within the obtained picture of world dynamics.

The analysis performed within the generalised effective (optical) potential method [21,23,25-30] shows that if one avoids its usual reduction to a version of perturbation theory [31,32], actually simplifying the problem down to a trivial one by simultaneously cutting all its essential dynamical links, then the solution can still be obtained, but in the form of *many* "locally" complete, and therefore mutually incompatible, *redundant* system configurations, or *realisations*, instead of only one such realisation in the case of invariable perturbative reduction of the conventional analysis. Since all realisations are equally real and "try to appear" under the influence of the driving system interaction, they should *permanently* replace each other in a *causally* unpredictable, *dynamically random* (probabilistic) sequence, which means also that each particular system realisation, representing an *arbitrary* system state (configuration), is intrinsically, dynamically *unstable* and will inevitably be *changed* for another one "chosen" by the system in a *causally random* fashion. That unceasing change of the whole system configuration determines the causally specified, universal *discreteness* of arbitrary system dynamics and related *irreversible* flow of intrinsic *time* driven by the *unreduced* interaction process itself. The natural discreteness of system dynamics emerges only together with, and therefore is inseparable from, the inherent dynamic randomness of the sequence of discrete realisation appearance that provides the ultimate, universal, reality-based and purely dynamic *source* (and the very *meaning*) of

---

[*]Knowing the incorruptible honesty of Max Planck's attitude to scientific results, one can be sure that his objection to the photon idea resulted from its obviously weak basis rather than any subjectively driven opposition to novelty. Being opposed to the idea, he accepted the original Einstein's paper for publication as the editor of the journal *Annalen der Physik*, in sharp contrast to today's self-interested manipulations of the dominating followers of the formal approach in science.

*randomness* in the world (unifying within it "unpredictability", or "chance", and "undecidability"). It is not surprising therefore that the universal science of complexity provides also the *dynamic* definition of *probability* and the method of its *a priori* calculation for *arbitrary* system.

The complex-dynamic, intrinsic discreteness, or *quantization*, is different from any formal, artificially imposed discreteness by its relation to internal *continuity* of real system dynamics, since while performing its permanent transitions between realisations, the system should pass by a particular, highly irregular "intermediate" state, or "main realisation", where the interaction components, closely entangled within each realisation, should transiently disentangle to a quasi-free state, in order to be again entangled into the next realisation configuration [21,23,28-30]. Therefore natural quantization of any real interaction process can be described as *qualitatively nonuniform*, highly uneven and *essentially nonlinear*, but *internally continuous*, rather than discontinuously punctuated/broken dynamics, even though both of them are opposed to *uniformly* continuous, or *unitary*, dynamics that inevitably results from the dynamically single-valued, effectively one-dimensional analysis of the conventional science.

The involved internal structure of any real interaction process in the form of unceasing dynamic entanglement-disentanglement is obtained simply as a result of truly rigorous, unreduced (universally nonperturbative) interaction description, as opposed to huge, qualitative reduction of usual perturbative analysis actually "killing" all but one system realisations and thus also intrinsic complexity of any system dynamics (where complexity itself is universally defined in the unreduced description as any growing function of the total number of observed system realisations, or related rate of their change, equal to zero for the unrealistic limit of only one realisation [21,23,26-30]). Apparent "stability" of external shape/dynamics of certain, "regular" kind of system, as if confirming the validity of conventional, single-valued (perturbative) modelling, is simply due to the fact that those particular systems have closely resembling, densely spaced realisations, so that their internal change can easily remain unnoticed (especially when not particularly sought for), but individually specified, *multiple* system realisations still always exist and *permanently change* each other "inside" their observed external envelope. One deals here with the limiting regime of *multivalued self-organisation* of complex dynamics, whereas the opposite limiting case of *uniform (global) chaos*, showing itself as visibly "irregular" and "nonlocal" behaviour, corresponds to sufficiently differing, broadly spaced system realisations. It is evident that this latter case of unreduced (multivalued) complex dynamics just corresponds very well to the observed "mysterious" properties of essentially quantum systems, including natural quantization, intrinsically probabilistic character and uncertainty. Note also that any version of the conventional "science of complexity", including usual "self-organisation"/"synergetics", "chaos", "criticality", "catastrophes", "multistability", etc., as well as related simplified, "model" imitations, computer simulations and empirically based speculations do not propose any equivalent of dynamic multivaluedness phenomenon and actually always fall within the same, dynamically single-valued, perturbative description of reality as the standard, "non-complex" science (see [23] for more details).

We see that the natural properties of unreduced complex behaviour of a system with interaction, rigorously derived within the unrestricted, reality-based analysis, provide at least qualitatively correct reproduction of "inexplicable" features of quantum behaviour, which should also be compared to the fact that dynamically complex behaviour is probably the unique possibility that has never been tried by usual theory as the origin of such properties, observed for both quantum and complex dynamics, as discreteness, duality (qualitative change of system state), nonlocality and randomness/unpredictability. Universal science of complexity shows [23] that those properties are unified manifestations of complex behaviour inevitably emerging (though with various visible magnitude) in *any* system of interacting components. Therefore, in order to obtain the well-specified causal extension of the standard, empirically based scheme of quantum mechanics, one needs to specify the particular system that gives rise to observed behaviour of elementary entities (particles, fields) and their structure as such. This requirement reveals the essential difference of the universal science of complexity from any canonical (positivistic, classifying) science version: the truly consistent, realistic understanding of the former implies *explicit derivation* of *all* observed entities and properties, in agreement with their emergence in natural dynamical processes, instead of formally fixing (postulating) the fact of their existence under the deceitful name of "theory confirmed by experiment". Every system, phenomenon, or level of reality is obtained from the unreduced *interaction* process and its development, instead of formal, axiomatic registration of its certain, artificially "chosen" and rigidly fixed results, whereas other ones remain "mysterious" and "inexplicable" (or even "incognizable", according to the Copenhagen type of "quantum mysteriology"). We call the unreduced description of the unified world complexity at its *lowest*, microscopic (or "quantum") levels *quantum field mechanics* [21-23,28-30].

## 2.2. Dynamic quantization and elementary field-particle emergence

The *simplest* possible system of interacting quantities that can indeed form the physical basis of the observed world is given by two *a priori* uniform, physically real media, or "protofields", homogeneously attracted to each other. Indeed, one cannot imagine yet simpler configuration of components that could give any further structure development (we refer to Occam's principle of parsimony), and at the same time we can show, using the unreduced interaction analysis of the universal science of complexity, that this simplest interaction configuration gives already autonomous emergence of elementary particles/fields, their interaction forces and all higher-level objects possessing the totality of experimentally observed properties, including the "essentially quantum" phenomena, now causally explained [21-23,28-30]. Such truly consistent description of the universal science of complexity involves only one "axiomatic" idea about a physical, rather than mathematical, origin that endows one of the interacting protofields with the electromagnetic (e/m) physical nature (because eventually it gives rise to the e/m entities and phenomena), while the second protofield is described as gravitational medium (because it gives rise to the gravitational interaction), in accord with universal and "extended" (like the protofields) occurrence of both e/m and gravitational phenomena. In fact, the actual protofield properties are causally specified later, by comparison of theory predictions with observations.

In other words, this "physical" postulate simply specifies the tangible "quality", or "type", of our world which is "made of" some light e/m matter coupled to a much more "heavy"/"inert" and "rigid"/"viscous" material of the gravitational "matrix"[*] (whereas other worlds could be obtained by the same, universal and causally derived, kind of development of their structure from interacting media of different types and numbers). We argue that any truly consistent and realistic world description should be based exclusively on that type of "material" postulate specifying eventually (after full development of the basic component interaction) "what kind of fruit this world is" with respect to other possible "fruits" on the "tree of Creation", as opposed to artificially imposed, greatly redundant number of abstract "axioms", "principles" and "fundamental laws of nature" of the conventional science that demonstrate only the fundamental *ignorance* of reality within *that* particular form of knowledge having nothing to do with Nature and knowledge about it *in general* and actually related, as we show within our unreduced, dynamically multivalued description, to evident limitations of its perturbative, dynamically single-valued, and thus *effectively one-dimensional* projection of dynamically multivalued reality. Extending canonical science up to the full richness of real, multivalued dynamics, we can reconstitute the Max Planck's *absolute* (= objective, consistently understood) *reality* starting from its lowest, "quantum" levels, now liberated from conventional inconsistency and related mystification, so decisively rejected by the "reluctant" father of the "new physics".

In particular, the introduced physical foundation of the world, in the form of two protofields, and especially its e/m component, can serve (in their quasi-free state) as the causal version of the classical Newtonian "aether", the necessary material, tangible basis of the universe, *actually* unifying it into a holistic, viable entity and so "proudly" rejected as "useless" by the triumphant 20th century positivism of the conventional relativity and other branches of "mathematical" physics. Since *any* real structure is obtained by development of the fundamental protofield interaction, the aether as such, in its pure form, cannot be directly observed (one cannot observe the basic "material" all the instruments are made of), but it emerges as a necessary material, unifying basis of the world with many particular manifestations, unexplainable without it [21-23,28-30]. Since every real entity, starting from space, time, and moving elementary particles, is produced from the aether by the driving protofield interaction, as it is shown in detail in quantum field mechanics, the naive mechanistic objection of the canonical science around the absence of "aethereal wind" for moving bodies (or permanence of the velocity of light) can rather prove than disprove the reality of *thus specified* physical aether.

---

[*]This asymmetry between the two protofield properties leads, in particular, to a definite "bias" in the *directly* perceived world structure, which is "displaced" towards the e/m constituent of the system.

Since one should not use any additional assumption or "model" within the truly fundamental approach, our description of the simplest configuration of two interacting protofields starts with the "existence equation", which simply expresses that configuration in a symbolic notation, without any refinement of interaction potential or other details that should be consistently derived in the theory:

$$\left[ h_\mathrm{e}(q) + V_\mathrm{eg}(q,\xi) + h_\mathrm{g}(\xi) \right] \Psi(q,\xi) = E \Psi(q,\xi), \qquad (1)$$

where $q$ and $\xi$ are *a priori* continuous (but in reality properly structured), physically real degrees of freedom of e/m and gravitational protofields/media, respectively, $h_\mathrm{e}(q)$ and $h_\mathrm{g}(q)$ are the corresponding "generalised Hamiltonians" describing the (unobservable) "free state dynamics" of protofields without interaction, $V_\mathrm{eg}(q,\xi)$ is the (eventually attractive) interaction potential, $\Psi(q,\xi)$ is the "state-function" describing the (developing) state of the compound system, and $E$ is the "eigenvalue" of the "generalised Hamiltonian" for this state (as the following analysis shows, "generalised Hamiltonian" is reduced to a measure of rigorously defined dynamic complexity [23]). In that broadly interpreted notation, the existence equation is equivalent to a general expression of practically any particular equation, but one should *not* reduce it to any special "model" at this stage. In particular, the fundamental protofield interaction potential $V_\mathrm{eg}(q,\xi)$ corresponds to the "unified interaction force" of the conventional field theory, and we later show (section 3) how this single starting potential gives rise to exactly four basic interaction forces with their causally specified origin and observed properties [21,23,28-30]. One can only gradually specify some properties of $V_\mathrm{eg}(q,\xi)$, in agreement with the observed results of its development. In a similar way, the starting equation cannot contain either time or space that should naturally *emerge*, as well as other real entities, as a result of unreduced interaction development rigorously ("exactly") described by our analysis.[*]

It is convenient to follow interaction development within the well-known optical, or effective, potential method extended to arbitrary system dynamics and its nonperturbative analysis [21,23,25-30]. It consists in the formally equivalent reformulation of the same problem, where all but one interaction participants are "excluded" from the main equation, even though they actually remain, but now in an "indirect" form of "effective potential" containing some "essential" parts of system dynamics. Thus, excluding the e/m protofield variables from eq. (1), one obtains the *effective existence equation* for a component, $\psi_0(\xi)$, of the total state-function $\Psi(q,\xi)$:

$$\left[ h_\mathrm{g}(\xi) + V_\mathrm{eff}(\xi;\eta) \right] \psi_0(\xi) = \eta \psi_0(\xi), \qquad (2)$$

where $\eta$ is the eigenvalue coinciding with $E$ from eq. (1), but actually corresponding to the emerging space-point "coordinate", and the operator of *effective (interaction) potential (EP)*, $V_\mathrm{eff}(\xi;\eta)$, is expressed through the free e/m protofield dynamics and unknown solutions for another, "truncated" problem with a reduced number of degrees of freedom (see [21,23,27-30] for more details).

---

[*]The variables $q$ and $\xi$ can correspond to "spatial" representation of the protofield matter, but that "aethereal", yet more fundamental level of dynamics and its "space" are not directly accessible in this world that emerges by development of configuration (1) and remains separated from the internal protofield dynamics by a large enough gap, in accord with intrinsic discreteness of complex dynamics deduced by the same analysis of unreduced interaction process.

An important feature of the "effective" problem formulation is the unreduced EP dependence on the eigensolutions to be found (explicitly shown by dependence on $\eta$ in eq. (2)), which introduces *essential nonlinearity* in *any* unreduced interaction process. The physical origin of this EP dependence and the resulting nonlinearity is in the essential dynamical links within the system explicitly entering the EP expression and describing the unreduced interaction process development. The latter is characterised by natural formation of *interaction loops*, since in a real system, where "everything interacts with everything", any system part acting upon another part also acts upon itself, by intermediation of other parts. Interaction loops are directly expressed by the "mathematical" loop between the two sides of eq. (2) maintained by their dependence on the same eigenvalue $\eta$.

As that complicated system of hierarchically entangled links creates an apparently "irresolvable" ("nonintegrable", or "nonseparable") problem, the conventional science, always trying to find some "simple", "finite" (or "closed") solution, applies a version of "perturbation theory" consisting in artificial reduction of essential dynamical links, which eliminates the EP dependence on the eigenvalues together with the related essential nonlinearity and dynamic complexity (i.e. simplicity kills complexity). As a result, perturbative analysis always gives a physically trivial solution describing practically the same system as the one without interaction, but with some weak perturbations of initial structure, proportional to the model potential magnitude (that *should* be small enough, even if it is not so in reality). Most important is the fact that such severely cut system dynamics cannot produce any qualitatively new structure in principle, but only handicapped "small variations" of the already given, dynamically uniform (unitary) system. This conclusion remains valid for those approaches of the dynamically single-valued, or *unitary*, science which are characterised as "nonperturbative" and "nonlinear" within that conventional paradigm (e. g. "nonperturbative field theories" or "solitons"): those "exact" solutions and all their perturbative modifications result from artificially inserted "curvature" and hand-made "involvement", but they always preserve their intrinsic separability, dynamic single-valuedness (unitarity) and therefore represent but the same fatal simplification of the dynamically multivalued, *essentially* nonlinear natural objects/phenomena. The whole conventional science, including scholar "science of complexity", "causal interpretations" of quantum mechanics and trickily "renormalised" field theories, is forced to artificially "introduce by hand", postulate the existence of its main entities, like space, time, elementary particles, fields and their "intrinsic" properties, remaining always limited, however, to their *abstract* imitations, which are simply "reconsidered" from a "different aspect". All of it is often reduced to simple adjustment of *any desired* number, $n_{th}$, of "free parameters (and entities) of the theory" to $n_{exp} \approx n_{th}$ "experimental results". That basically ambiguous, purely combinatorial fitting of perturbative imitation results to *subjectively selected* observations is called "excellent agreement between theory and experiment" providing "decisive support" for the theory. As the number and choice of "parameters" and abstract entities in various "theories" can be varied practically infinitely (including any desired number of world's dimensions or even type of its "logic"), that kind of "scientific progress" inevitably leads to the current state of "exact" science where any nontrivial phenomenon is equally "successfully" explained within many such competing, purely abstract theories, whereas the actual physical nature of participating entities and occurring processes remains "mysterious".

If one avoids any perturbative reduction of the full EP formalism, then it appears that the solution to a problem, in its "effective" formulation of eq. (2), can still be found, but in a *multiple, redundant* number of "versions", each of them exhaustively characterising system state and therefore called "realisation". Being rigorously derived by the unreduced EP analysis in its both "geometric" and "algebraic" version [21,23,25-30], the dynamic redundance feature originates in essential nonlinearity of any real interaction process, increasing the power of the characteristic equation for the eigenvalue, and has a transparent *physical interpretation*. If one imagines that the homogeneous initial configuration of interacting protofields is modified by a small local fluctuation of increased density of one of the fields, then it will produce a density increase of another protofield around the same location, which will amplify the fluctuation of the first protofield, etc., until the emergence, in a "catastrophic", avalanche-like fashion, of a highly increased density of dynamically entangled protofields that actually constitutes the elementary particle "core". However, the system dynamics cannot stop at the stage of maximum protofield compression (determined by their finite compressibility), since the participating protofield parts continue to be attracted to their other parts, which creates the same type of instability, and the system catastrophically disentangles towards a quasi-free state, before being entangled around a new centre of interaction-driven "reduction" (self-amplified dynamical squeeze). Note that the position of the next reduction centre is "chosen" by the system in a dynamically random fashion (due to unpredictability of small fluctuations), which corresponds to the *ultimate, causal source of randomness* in the system of interacting protofields (or any other real system) that inevitably produces a *dynamically redundant* number of realisations (which are different versions of interaction component entanglement).

The dynamic multivaluedness phenomenon is also confirmed by the following simple argument. If the protofield interaction is described as attraction between $N$ "points", or "elements", of each protofield, then the unreduced interaction of "everything with everything" produces $N^2$ versions of those elements entanglement. However, there are always only the same $N$ places for interaction results, since the number of places (or "volume") occupied first by interaction partners and then by interaction products cannot change without introducing a "bad", non-dynamic arbitrariness in the emerging structure (this "evident" rule is a particular manifestation of the universal symmetry/conservation of complexity [23]). The $N$-fold redundancy of interaction results becomes evident and leads inevitably to their unceasing change in a causally random order. Those transparent confirmations of the dynamic redundance phenomenon, in addition to its rigorous derivation within the universal, rigorous analysis, demonstrate once more the advantages of the "absolute reality" paradigm over blind manipulations of conventional, "mathematical" physics dominated by the evidently inconsistent and severely limited paradigm of dynamic single-valuedness.

The dynamically redundant entanglement does not stop at the first level of system splitting into incompatible realisations, but continues in a series of levels of ever finer splitting into internally entangled realisations that form the *dynamical fractal* of a problem. It is the *dynamically probabilistic*, causally complete extension of the ordinary, dynamically single-valued fractals, obtained due to "non-separability" (non-integrability) of the problem reflected in the relation of obtained solutions of the effective existence equation, eq. (2), to solutions of an auxiliary, truncated problem [21,23,25-30]. Dynamically probabilistic fractal describes the irregular and changing internal structure of the emerging new entity (the elementary particle in the case of interacting protofields, see below).

The above picture reveals the basic mechanism of dynamic multivaluedness and entanglement, which is the intrinsic, *dynamic instability* of *any* real interaction process directly reflected by the interaction loop formation and expressed by the essential nonlinearity of the "effective", actually more adequate problem formulation of eq. (2). That dynamic instability is a permanently present, inherent property of unreduced interaction corresponding to unceasing realisation change and taking the form, in the case of interacting protofields, of permanently locally squeezing and extending (entangling and disentangling) protofields, which choose each time a (neighbouring) centre of squeeze at random.[*] The same process can alternatively be described as chaotic wandering of the dynamically squeezed, "corpuscular" protofield state called *virtual soliton* [21,23,28-30] in order to distinguish it from ordinary solitons, being totally *regular* solutions of particular, very "specifically" structured equations. Of course, the observed virtual soliton "jumps" always pass by the intermediate disentangled state of transiently "free" protofields that ensures internal continuity of qualitatively nonuniform dynamics. This particular, "main" system realisation relating all other, "regular" realisations (different virtual soliton positions) in one, holistic dynamics is none other than the causal, physically real version of the system *wavefunction* (it has now the absolutely universal meaning and extends also the notion of "distribution function" for "macroscopic", classical systems) [21-23,28-30].

---

[*]Dynamic instability against squeeze has also a *rotational* component giving rise to a highly nonlinear "whirlwind" of reduction-extension process and providing causal explanation of the intrinsic property of *spin* [21,23,28-30]. That inseparable unification between "rectilinear" reduction-extension and rotational spin motion in the same protofield interaction dynamics, uniquely provided by the essential nonlinearity of its unreduced version, permits one to understand the universal emergence of the *same* Planck's constant in *both* nonrelativistic *and* relativistic quantum phenomena, in discrete energy level structure and quantized values of spin [23]. In this paper we shall not explicitly analyse the spinning component of complex particle dynamics considering that it constitutes an integral part of the dynamically fractal, essentially nonlinear motion within the reduction-extension cycles of the elementary particle.

Note that all these results are obtained within the rigorous analysis of unreduced EP equations [21,23,28-30]. In particular, the eigenvalue $\eta$ describes the (emerging) "coordinates" of the protofield reduction centre, and the redundant number of its values reflects the dynamically chaotic wandering of this centre (virtual soliton) in the course of unceasing reduction-extension cycles of the interacting protofield dynamics. That process of essentially nonlinear pulsation in the *a priori* homogeneous system of interacting protofields, also called *quantum beat*, can be interpreted as *both* causal version of *elementary particle* (exemplified by the electron as the simplest quantum beat species) *and* dynamic emergence of *physically real space and time*, which provides the realistic, well-specified *meaning* for those most fundamental, "embedding" entities of physics that have always been introduced only as purely mathematical "coordinates" in conventional science (those *abstract* space and time are, in addition, inseparably "mixed" and "deformed" by a *real* physical mass, according to the main postulate and idea of the canonical general relativity).

The physically real *space* emerges as the discrete, highly inhomogeneous structure of interacting protofields (variables $q$ and $\xi$ in eqs. (1), (2)) dynamically "woven" (entangled) into fractally structured tissue as it is determined by the complete, nonperturbative solution of "effective" problem, eq. (2) (see refs. [21,23,28-30] for the detailed expressions). The smallest structural element, or physical "point" of that real space is given by the same virtual-soliton, "corpuscular" state of dynamically squeezed, entangled protofields that provides the "particle-like", localised aspect of the simplest material object of the world, the elementary particle. The related structural elements of space are provided by discrete jumps of the virtual soliton, which form the typical length element equal to the Compton wavelength and further structures, such as de Broglie wavelength of a moving particle [21-23,28-30]. In that way quantum field mechanics cancels the fundamental gap between the notions of "embedding" ("mathematical") space and material "object" embedded in it, inevitably reappearing in canonical science despite any artificially inserted "vacuum fluctuations", since we show that the dynamically changing, tangible tissue of real space is *directly formed* in the *same* complex-dynamical process of protofield *interaction*, or "quantum beat", that represents the physical essence of elementary particle described as essentially nonlinear *process* of permanent change between its two qualitatively different states, the localised, corpuscular state of virtual soliton and extended, undular state of physically real wavefunction. The latter property explains the intrinsic, dynamic duality of any such particle-*process*, or *field-particle*, including its behaviour in interactions ("quantum measurement", "wave reduction", etc.) [21-23,27-30]. The same quantum beat process gives rise to causal *time* that, contrary to space, is *not* a material, "tangible" entity forming a "dimension", but rather a sign of intrinsic inhomogeneity of the driving interaction process marked by well-specified *events* of reduction-extension cycles that play the role of the most fundamental, omnipresent "pendulum" of the universe. This physical difference between space and time explains why time, contrary to space, unceasingly and therefore "irreversibly" flows and shows that real space and time *cannot* be "mixed" in a unified "spacetime manifold". Both causal space and time emerge together with *dynamic randomness* of the quantum beat dynamics and could *not* exist without the *causally probabilistic* choice of each next reduction centre (or "physical space point").

We come now to the *complex-dynamical origin of discreteness* of all fundamental entities and their properties, which is the central point in the present discussion of quantum field mechanics commemorating the centenary of the first, "reluctant", but explicit emergence of "quantum discreteness" in Max Planck's work. We can now specify the detailed physical origin of this intrinsic discreteness as being due to the irreducibly *complex* dynamics of the underlying *interaction process* leading to (redundant) *multivaluedness* of interaction results, which explains also why the causal origin of quantization *cannot* be obtained within the dynamically single-valued, "non-interactional" (perturbative) approach of the canonical science, irrespective of its particular version or mathematical tools applied.

Discreteness of the most fundamental emerging entities, causal space and time, is obtained, as we have seen above, as a result of holistic, *essentially nonlinear* character of the quantum beat process of protofield interaction. The avalanche-like dynamical squeeze of a portion of protofield "material" and the following reverse process, involving certain its finite "amount" (depending on the interaction magnitude), can happen only during a finite, "discrete" time period, $\tau_C$ (the period of one quantum beat cycle), which is actually the smallest real time interval (its natural unit) emerging together with time itself.* It is determined by the EP amplitude for the elementary particle (it can be the electron for a majority of "ordinary" processes) equal to its rest energy (see below). Correspondingly, the elementary virtual soliton jump, determining the effective unit of causal space, is equal to the finite value of (generalised) Compton wavelength, $\lambda_C = \tau_C c$, since every (e/m) protofield perturbation propagates with the velocity of light *c*.

Space structure discreteness is obtained also as discrete values of reduction centre coordinates, $\eta$, found from the effective existence equation, eq. (2) [21,23,28-30]. That realisation distribution discreteness is a universal property of complex interaction dynamics resulting simply from its unreduced character. Indeed, in a system where "everything interacts with everything" each local motion gives rise to a whole series of mutually influencing motions that will influence the starting motion ("interaction loop"), and therefore arbitrary small motions are impossible (which reflects intrinsic instability of unreduced interaction mentioned above). As a result, interaction process development finally produces a certain number of *actually possible* collective motions, or system realisations, playing the role of slightly more stable (but still basically unstable) "turning points" of complex interaction dynamics, which are the necessary complement to intrinsic instability of a finite system. Such discrete structure formation of causally emerging space and time happens even in the *a priori* totally homogeneous configuration of interacting protofields and is therefore a truly "emergent" phenomenon which, nevertheless, is causally determined in all the resulting structure details by the protofield interaction magnitude (and their internal mechanical properties). It is that, truly fundamental nature of unreduced interaction properties described by the extended EP formalism that can uniquely provide the "definite", truly causal solution of "quantum enigma" accumulated by hopelessly reductive single-valued science that can see only an effectively *one-dimensional projection* of *multivalued* dynamics, looking as "mysterious" as every strongly reduced projection (cf. the "shadows" metaphor with respect to conventional science paradigm in the Roger Penrose book [13]).

Dynamic discreteness of the causally emerging structure of space and time gives rise to discreteness of mechanical *action* appearing as a natural, universal *measure of dynamic complexity* itself, as it is expressed by the complete set of permanently changing system realisations. Indeed, the most general measure of that reality-based complexity should be proportional to the natural, *spatial* measure, $\Delta x$, of the main structural element (elementary space/length unit) produced during a quantum beat (realisation change) cycle (thus $\Delta x = \lambda_C$ for the electronic quantum beat process and the whole realm of related quantum phenomena). For the same reason it should also be proportional to the time unit, $\Delta t = \tau_C$, dynamically emerging in the same realisation change (quantum beat) process. Those two aspects, spatial and temporal, provide an exhaustive description (measure) of the elementary "quantum of complexity", $\Delta\mathcal{A}$, emerging in one quantum beat cycle in the form of space structure element. Since the two aspects should be independent from one another (to allow for their observed independent variation), we conclude that $\Delta\mathcal{A} = p\Delta x - E\Delta t$, where $p$ and $-E$ are coefficients. As this relation should be compatible with similar ones, in both quantum and classical mechanics (recall that our analysis remains valid for any system with interaction), we conclude that our "complexity measure" $\mathcal{A}$ is none other than action, $p$ is momentum, $E$ is energy, and $|\Delta\mathcal{A}| = h$ is Planck's constant for the level of "quantum" phenomena. It is evident that as the quantum beat process performs its unceasing pulsation, action increments for individual cycles add up, so that for $n$ cycles one will have $|\Delta\mathcal{A}| = nh$ (it can be shown [23] that $\Delta\mathcal{A}$ is always negative and the minimum increment $\Delta\mathcal{A} = -h$).

---

*The number of elementary particle species emerging in the protofield interaction process is also greater than unity, each of them corresponding to a certain EP realisation in eq. (2) with the particular potential well depth and width. Correspondingly, one obtains different values for the quantum beat period $\tau_C$ determining the emerging time unit. The smallest of those periods should be considered as the most fundamental time interval of the universe (it corresponds to the heaviest elementary particle and can be of the order of $10^{-27}$ s [21,23,28-30]). In practice the respective levels of dynamics can be rather well separated, so that for a particular broad class of phenomena one deals with the respective, *effectively* smallest, value of $\tau_C$ that can be much greater than the ultimate, "universal" time unit. Thus, for generic "quantum" phenomena the level of *electron* quantum beat dynamics is the "operational" one, which gives $\tau_C \sim 10^{-20}$ s.

That causally derived new meaning of apparently "familiar" quantity of action as the universal integral measure of dynamic complexity [23] actually involves a number of qualitatively new findings. Most important for us here is the causal, *dynamic* origin of the quantum of action, or Planck's constant, $h$, that consists in the fundamental discreteness of *complex interaction dynamics* showing itself, in this case, at a group of its *lowest* sublevels observed as "quantum" phenomena. The central role of *action* quantum in description of this natural quantization process is due to the genuine, causal meaning of action as the basic integral *measure of (unreduced) complexity* emerging as the simplest unification of two main "products" of complex dynamics, causal time and space structure. Both this role of action and the involvement of complex dynamics of unreduced, omnipresent *interaction* in "quantum" phenomena totally escape the conventional quantum theory operating within a severely reduced, effectively one-dimensional projection of reality, which inevitably leads to "mysteries", ruptures and practical difficulties in the unified knowledge development. The result of unreduced, truly consistent description of the hierarchy of complex world dynamics shows that the minimal (absolute) change of action is $h$, which corresponds to a well-specified simplest motion element, quantum beat cycle, that still appears to be a quite "entangled" (also in the direct sense), essentially nonlinear, fractally structured and dynamically probabilistic result of *a priory* structureless interaction between two protofields. Since our analysis is actually universal, it shows also that action, together with space, time and other emerging entities and properties, changes discretely for any real system with interaction, though at higher, "non-quantum" levels of complexity the respective "quanta of action" $|\Delta\mathcal{A}|$ certainly exceed $h$ and need not be as permanent as $h$ (they are always determined by the characteristic spatial and temporal units, $\Delta x$ and $\Delta t$, for each particular system, which in different cases may or may not correspond to characteristic dimensions of externally observed, often only "statistically averaged" elements).

It is important to note the direct involvement of *dynamic randomness* of the quantum beat dynamics with its intrinsic discreteness giving rise to $h$, since this unified, complex-dynamic origin of quantization and randomness is in perfect correlation with Planck's original introduction of quanta within a thermodynamical problem analysis and especially with his persistent opinion, often considered by later "mathematical" physics as a "stubborn conservatism" and rejected as another "useless conviction", that those "thermodynamical" laws should have a fundamental (= "dynamical") basis instead of being a purely "statistical" consequence of a large number of system components. And although Max Planck was finally forced to formally acknowledge the apparent "victory" of the "statistical", purely "mathematical" (and fundamentally deficient) interpretation of his favourite "second law", the results of the universal science of complexity confirm now the rightness of his true, original attitude and reveal the purely dynamic, fundamental origin of randomness and related entropy growth in *any* real process, at the quantum level and beyond [23]. We can clearly see also why the dynamic origin of randomness, discreteness and other related manifestations of the unreduced complexity could not be discovered in principle within the conventional, basically single-valued science. The deep attitude of Max Planck, Louis de Broglie, or Erwin Schrödinger towards the unknown was always characterised by the search for a truly consistent, realistic explanation of a "mystery", in sharp contrast to "fundamental mysticism" within conventional "exact" knowledge inevitably leading to its today's crisis.

The fact that the seemingly "well-known" mechanical action function has that unexpected, complex-dynamical meaning is also important in itself. It clearly demonstrates the incompleteness of conventional, "well-established" picture of dynamics even at its "perfectly understood", classical levels, since it appears now that the deceptively "linear" mechanical action, entering in the conventional scheme through "infinitesimal" increments of smoothly varying space and time hides within it the essentially nonlinear and qualitatively nonuniform, catastrophically pulsating and chaotically "jumping" realisation change process always occurring through unceasing "quantum" reductions and extensions of the generalised, causal wavefunction (or "distribution function") of a system. Knowing now that action is the most fundamental integral expression of the *universal* dynamic complexity, one can better understand a special, "chosen" role the action function and related formalism play in various domains of physics, despite other, formally "equivalent" and often apparently more "natural" ways of conventional analysis (e. g. in classical mechanics). Moreover, the discovered realistic implication of action in the causally extended quantum mechanics and beyond leads to its still more profound interpretation as a really existing, permanently irregularly changing "field" of realisation probability distribution providing the causal, physically real analogue of conventional abstract "phase space", which is now equivalent to the causal wavefunction and related to it by dynamically extended "quantization rules" [22,23,28-30]. The latter provide causal relation between both realistically interpreted entities, action and the wavefunction, determined by the same fundamental action change, $\Delta \mathcal{A} = -h$, during each *real* "quantization act", the quantum beat cycle. That causal quantization mechanism allows for causally complete extension of other purely abstract ideas of the conventional theory, such as formal "creation" and "annihilation" of particles represented in reality by "corpuscular" states/realisations which are permanently "created" and "annihilated" within the same dynamically continuous protofield interaction process (in each quantum beat cycle), without any "help" of purely abstract, artificially introduced "operators" (see ref. [23] for more details).

Causal extension of action and its quantization inevitably involves also complex-dynamical extension of the quantities of *momentum* and *energy* related to action by the (dynamically) discrete analogues of partial derivatives:

$$p = \frac{\Delta \mathcal{A}}{\Delta x}\Big|_{t=\text{const}} = \frac{|\Delta \mathcal{A}|}{\lambda} = \frac{h}{\lambda} \; , \quad E = -\frac{\Delta \mathcal{A}}{\Delta t}\Big|_{x=\text{const}} = \frac{h}{\tau} = h\nu \; , \tag{3}$$

where, for the general case of moving field-particle, $\lambda \equiv (\Delta x)|_{t=\text{const}}$ is the emerging "quantum of space" (or "de Broglie wavelength" [21,23,28-30]), a major directly measurable (regular) space inhomogeneity of a quantum system with complexity-energy $E$ ($>E_0$, the rest energy), $\tau \equiv (\Delta t)|_{x=\text{const}}$ is the system quantum beat period measured at a fixed space point, and $\nu = 1/\tau$ is the corresponding frequency. We see that momentum and energy are universal *differential* measures of the same dynamic complexity, corresponding to its integral measure by action: momentum characterises the spatial rate of complex-dynamical structure emergence, while energy characterises the temporal rate of the same process. The functional relation between the two quantities, $p = p(E)$, specifies their contributions to particular system complexity, i.e. shows whether the latter is produced as a spatially more complex ("short wave-length"), but slowly emerging structure, or as a spatially less involved ("long wave-length"), but quickly emerging structure. For example, in the case of sufficiently uniform emergence of fine structure, giving the generalised "trajectorial" or "classical" type of dynamics, one has $p = (E/c^2)v = mv$, where $v = \Delta x/\Delta t$ is the velocity of structure emergence, or "motion". Therefore the $p(E)$ dependence, sometimes called "dispersion relation", actually specifies the dynamical motion regime (behaviour), or simply (motion) *dynamics*, for a given type of system. It is this dynamic relation between the spatial ("tangible", "textural", "material") and temporal ("emergent", "event-like", non-material) aspects of system complexity, specifying its properties, which is imitated in the conventional-science description of direct relation between motion, space and time (or "relativity") by a formally postulated "mixture" between space and time, within a single abstract "manifold" and "geometry/topology".

Note the radical change in energy interpretation implied by the obtained picture and eqs. (3): instead of conventional ambiguous, formally defined characteristic of "swiftness of (certain type of) motion" (the latter being also "defined" only "intuitively"), the causally extended, "real" energy is consistently obtained as the temporal rate of complex-dynamic structure emergence, where causal time appears in the form of dynamically discrete sequence of reduction/extension (realisation change) events, as described above. It is important that the causally specified energy (now universally defined for *any* real process) characterises a highly nonuniform, essentially nonlinear, dynamically random realisation change that can be found within each, including externally "smooth", motion. This important result shows *why* energy is "equivalent to mass" (a "famous", but actually only formally postulated conclusion of the conventional relativity) and simultaneously provides the universal causal interpretation of *mass* as the same temporal rate (divided by $c^2$) of *dynamically random* realisation change (virtual soliton wandering), which consistently explains the main property of *inertia*. In its turn, *motion* itself is universally defined now as a system state with energy-complexity exceeding its *minimum* value called *rest energy* and characterising the *state of rest* (always existing and well-defined) with *maximum* dynamic randomness of realisation change process (realisation probability distribution is most uniform and any "global", ordering tendency is absent) [21,23,28-30]. We are not limited now to empirically based, inconsistent notion of "reference frame" and related ambiguous "position measurements" in order to conclude that the system of arbitrary kind and complexity is at rest or in a state of (global) motion. This is another step towards the objectively understood, absolute reality of Max Planck.

In that way it becomes clear why quantization of action-complexity at the level of quantum phenomena, expressed by Planck's constant and reflecting intrinsic discreteness of complex quantum beat dynamics, incorporates (now dynamically explained) quantization of other emerging entities, such as space, time and energy levels within a bound system (see ref. [23] for more details). It remains for us to understand the incredible universality of the quantum of action, appearing in so many phenomena of vastly different scales and origin. Various elementary entities and their interaction regimes are interpreted within quantum field mechanics as different realisations of the fundamental interaction between two protofields (and their further dynamical splitting, by the same mechanism, into a hierarchy of higher realisation sublevels). Each realisation possesses well-defined EP characteristics and thus gives rise to (quasi-)permanent quantization properties. One needs to understand, however, why different EP realisations (for example for different elementary particle species) give rise to the same basic quantum of action, $h$ (unless it is a mere consequence of the formal play with "free parameters of the theory", which seems not to be the case, at least for an important part of the whole diversity of $h$ emergence). This fact can be explained, within the causal picture of quantum field mechanics, if we make a realistic assumption that the "deformation energy" of protofields forming EP well is proportional to the "volume" of the well cross-section, since the latter is just determined by the corresponding action quantity, $\Delta \mathcal{A} = (\Delta p)(\Delta x)$ (where $\Delta p$ accounts for the EP well depth and $\Delta x$ for its width). As the "deformation energy" should be permanent for a given fundamental interaction strength, it appears that we may have, for different EP realisations, either a shallow, but wide pit, or deep, but narrow hole as EP well configuration, but the quantity that remains constant for all of them is $\Delta \mathcal{A} = h$. Although this explanation is a tentative one and may need further refinement (it cannot, of course, be verified by a direct measurement), it clearly demonstrates a realistic principle of world construction that provides unified causal explanations for various, otherwise quite "mysterious" properties of the observed quantum behaviour. In particular, we can still better see now why exactly it is the fundamentally fixed action (and not, say, energy) quantum that determines the *unified* diversity of the naturally emerging world structure. Causally interpreted action quantum and its permanence directly reflect the necessary *physical*, "material" unity of the complex world dynamics that leads to the universal "order of the world", rigorously specified in the form of "universal symmetry/conservation of complexity" [23]. The detailed theory of quantum field mechanics shows [21,23,28-30] that the same is true for other "universal constants", which simply express well specified dynamical properties of the interacting protofield system. Such physically transparent, rigorous derivation of the observed fundamental properties of absolute reality from its objectively specified, irreducible starting configuration is certainly very close to Max Planck's vision of the desired kind of scientific knowledge [2,6,8].

### 2.3. Classical behaviour emergence and causally extended relativity

Having explained the origin and permanence of $h$ in the world of quantum phenomena, one should now also consistently explain its "sudden" *disappearance* in "classical" (trajectorial, localised) behaviour, that is in our "usual", macroscopic world, which cannot be separated, of course, from causal understanding of the fundamental origin of "classicality" and its relation to "quantum", "nonlocal" behaviour, constituting another "quantum mystery" of usual "exact" science. We argue that classical "localisation" of naturally "delocalised" quantum beat dynamics emerges as a next higher level of complexity, when elementary field-particles, causally interpreted as quantum beat processes, start interacting among them (through their common forming media, but mainly the e/m protofield in this case) and form, in particular, elementary *bound systems* (like atoms). Indeed, if such bound system can exist as a reasonably stable, well-defined entity, it means that its elementary constituents (for example, electron and proton in the hydrogen atom) that naturally *continue* their intrinsic *random* walk cannot easily perform a larger advance (exceeding the equilibrium bound state size) *in the same direction*, since for this they would need to perform a large series of *correlated*, but *independently random* quantum jumps (when *all* of them advance in almost one direction, but *each* of them always chooses it *at random* among *all* possible directions). It means that the *probability* of truly delocalised, chaotic wandering of a bound system *as a whole* within its wave that just determines the "essentially quantum" type of object behaviour is very low (it is "exponentially small" outside the system equilibrium size). The true reason for localisation of elementary (and larger) bound system trajectory is related, paradoxically, to unreduced *dynamic randomness* of "nonlocal" quantum behaviour of *each* of its constituents implying also intrinsic mutual *independence* of constituent quantum beat processes always determined by a more fundamental (and therefore generally much stronger) protofield interaction process. Therefore, although the universal "event arena" always remains the same (the interacting protofields couple perceived from its e/m side), the $h$-determined, properly "quantum" dynamics is now safely "hidden", with all its "nonlocal jumps" and *associated* "waves", *within* the elementary bound systems (atoms, nuclei) that form progressively the whole diversity of higher-level, "macroscopic" structures. The motion of an individual bound system as a whole will also be quantized, of course, due to universal complex-dynamical quantization of any real interaction process, but now the emerging "quanta" of action, space, time, etc., as well as the corresponding "uncertainly" of system trajectory, will be determined by the new, higher-level dynamics, without any direct relation to $h$. Higher-level quanta will have much larger, and growing, diversity at every higher level of complex world dynamics (in agreement with the unreduced complexity definition), while their typical manifestation will be reduced to dynamical chaos, more or less "localised" around the average "trajectory" (see [22,23] for further details). It is important that now any emerging quantum is *actually* divisible into a fractal hierarchy of *observable* finer quanta, so that the system is not limited to big jumps, as it occurs at the lowest, "quantum" levels of complexity.

Note that due to unique properties of unreduced dynamic complexity quantum field mechanics provides the purely dynamic, *intrinsic* origin of classicality that does not depend on any changing "influence of environment" or ambiguous, formally imposed "decoherence" of "state vectors" from an abstract "space", inevitably evoked by conventional "explanations" of classical behaviour emergence (even though external influences may play a quantitative role, different for each particular system). We can understand also, within our complex-dynamical picture, why certain *interacting* bound systems can "reconstitute" the lost "quantum" properties, despite their sometimes quite large mass, complicated structure, and interaction with the environment implying strong "decoherence" (see ref. [29] for the detailed explanation of existing experimental results).

After having causally obtained the classical type of behaviour as a next higher level of the same dynamic complexity that provides the natural origin of quanta at its lowest level, one may be interested in possible inclusion into that unified picture of another "corner stone" of the "new physics", also much supported by Max Planck, the theory of "relativity", or dependence of measured space and time scales on the relative system motion (and the surrounding gravitational field). It is a well-known fact that the canonical interpretation of corresponding relations of Lorentz and Poincaré proposed by Einstein is only mechanically "added" to the first, "quantum" branch of the "new physics", in a similar form of a number of artificially imposed, abstract "principles" and accompanying mathematical "guesses", so that the two "great theories" remain basically separated among them, and even qualitatively contradicting to one another (despite their formal "joining" in certain equations), while each of them is based on a number of independent and "inexplicable" postulates. Problems arising from that inconsistency of scholar fundamental physics are so important that "true" unification between quanta and relativity is often considered as the "last big quest" of the (otherwise "perfect") modern physics [33]. However, it tries to "organise" for a suitable marriage of the two opposed partners using the same, purely mathematical, single-valued, "fixed-structure" paradigm in the form of modern "super-abstract" branches of canonical "field theory" ("strings", "branes", "M-theory", etc.). The result shows that instead of expected "magic" of "unreasonably efficient" mathematics, one obtains only unreasonably redundant number of ambiguous, abstract guesses involving, in particular, various numbers of "hidden dimensions" and strangely "invisible" entities, where each scheme "could be true", but actually cannot consistently explain already the "old good" quantum and relativistic properties, which preserve their original "mysterious" status fixed by "postulates" and "principles" (now greatly increased in number).

Contrary to that hopeless search in the ever denser jungle of incomprehensible abstractions, quantum field mechanics provides *intrinsic*, dynamic and reality-based unification of "quantum" and "relativistic" properties within the *same* unreduced analysis of apparently simple system of two uniformly interacting protofields. Quantum beat dynamics resulting from that interaction and described above contains both "quantum" and "relativistic" features from the very beginning, unified as different aspects of the same complex behaviour naturally emerging also at all higher levels of complexity (due to universality of the unreduced analysis) [23]. The basic origin of space and time dependence on the system motion is in the very fact, specified by quantum field mechanics (see the previous section) that space, time, and motion *explicitly* emerge *all together* in one and the same process of chaotic realisation change and therefore cannot be independent of each other: a change of system's state of motion will inevitably influence the properties (observed rate) of related intrinsic time with respect to that produced by unchanged, "reference" state of motion (or the state of rest). It is evident, therefore, that such explicit indication of the *fundamental physical origin* of "relativistically modified" entities (space, time, mass-energy, etc.) and that of motion (and gravitation) is absolutely necessary for the consistent, truly causal understanding of their "relativity" within *any* approach, which explains the failure of canonical science picture that tries to "circumvent" the "unnecessary realism" and inserts space and time as "symbolic" elements in its formal constructions providing them with "suitable", postulated "rules" (like "curved space-time geometry" of the conventional general relativity).

The basic effect of special relativity, "relativistic time retardation", measured within a moving system with respect to that measured in the state of rest, is derived in the universal science of complexity [21,23,28-30] in the form of more intense quantum beat dynamics within the moving field-particle (or any larger system), while the remaining "random deviations" from that "global" tendency, just determining the *actually produced* (and therefore *measured*) time flow, inevitably become less frequent (with respect to the state of rest). Elementary derivation based on the above relations between causal time, mass-energy, and action (see eqs. (3)) reproduce, of course, the canonical Lorentz-Poincaré expressions (see refs. [21,23,28-30] for the details), but now they are rigorously obtained from the unreduced interaction analysis without any artificially imposed "principle of relativity" (we always use only our "mild" assumption about the physical protofield nature, see section 2.2). The velocity of light naturally appears as velocity of perturbation propagation in the e/m medium, while its famous independence of emitting body motion is due to the deduced different time flow rate (frequency) for the moving body. The same analysis provides causal version of famous mass-energy equivalence (already mentioned above), the "relativistic mass increase", and other related effects of special relativity obtained in its canonical version by formal adjustment of postulated mathematical constructions like Lagrangian to the known final expressions of Lorentz and Poincaré (using a number of additional assumptions). It is evident that intrinsic realism of the causally extended interpretation of "relativistic" effects and "universal constants" corresponds much better to the absolute reality paradigm of Max Planck who always emphasised, again contrary to the mainstream attitude, the "absolute" aspects of "relativity".

We can only mention here the equally physically transparent origin of gravity and related effects of "general relativity" within the same, complex quantum beat dynamics essentially involving now the gravitational protofield as the second participant of the driving interaction process (see refs. [23,28-30] for further details). Similar to observed e/m interactions transmitted, in the form of (photonic) perturbations, through the e/m protofield between two particles (quantum beat processes), the *intrinsically* universal gravitation is due to indirect exchange between the same processes occurring through the common gravitational medium/protofield, which incorporates the inherent, causally extended "equivalence between inertial and gravitational masses". Gravitational field of a particle (body) is interpreted as increased average (but internally *quantized*) tension/density of the gravitational protofield leading to modification of any other quantum beat frequency (depending on the distance between them), which explains the origin of *both* gravitational attraction (gravitational mass) *and* the effect of "time retardation in the field of gravity". The causally derived expression for the magnitude of the effect agrees with the conventional one, but it does not depend on any ambiguous "curvature" of abstract and formally "mixed" space and time (both space and time remain, of course, "flat", though internally inhomogeneous distributions of the coupled protofields density and quantum beat frequency).

The obtained world picture is liberated from several "difficult" problems of conventional relativity due to its artificially "curved" universe, appearing especially in cosmology. It is important that the causally complete picture of quantum field mechanics involves intrinsic, "inbred" unification of entities and properties, including causally extended "quantum mechanics", "special relativity", "general relativity", all observed "material" entities (particles and fields), their "intrinsic" properties (such as mass, electric charge and spin), and exactly four "fundamental forces of nature" [21,23,28-30] (see also section 3). Quantum effects, transition to "classicality" and other step-like phenomena reflect the dynamically discrete character of the unreduced complex dynamics, while phenomena like relativity account rather for its internal continuity appearing "between" the discrete states and transitions. It becomes clear also why there is no sense to look for any of the above intrinsic unifications within the dynamically single-valued, inevitably ruptured constructions of the canonical theory.

### 3. Theory confirmation by experiment and particular problem solution

Internal consistency and unity of quantum field mechanics (and the universal science of complexity in general) are supported by its successful application to various particular problem solution, including the majority of stagnating, "unsolvable" problems of usual fundamental physics, and we can give here the following outline of obtained solutions (see refs. [21,23,26-30] for more details). They can be compared to purely abstract, actually always postulated "explanations" of ruptured conventional theories, including the latest "elegant" imitations of string theory that operates with any desirable number of dimensions with adjustable parameters and over-simplified, abstract images of real entities.

**(1)** *Unified causal explanation of quantum and relativistic particle properties*. First of all, we cannot avoid mentioning again the obtained *causal, intrinsically unified* explanation of the physical origin and observed properties of elementary particles, fields and their interactions, always remaining "mysterious" and mutually separated within the conventional theory and any its technical modification.

As we have seen above, not only we *rigorously* derive the detailed internal structure of space, time and their manifestations in the form of naturally emerging particles and fields, but those entities are provided with the intrinsically "incorporated" and *dynamically unified* fundamental properties (including wave-particle duality, indeterminacy, uncertainty, quantization and relativity) which are permanently *observed* in experiment, but can be only formally postulated, *without any* consistent explanation, in canonical, *irreducibly separated* versions of quantum mechanics, relativity, field theory, and particle physics. Among various simultaneous unifications, one should emphasize the intrinsic unity between quantization and relativity which both emerge as different aspects of the same complex-dynamical interaction process within any real entity, so that e. g. "relativistic time retardation" *results from* the *quantized* and *dynamically random* origin of time [23,28-30]. We can understand also why consistent unification of the two externally "conflicting" theories, considered as "quantum theory's last challenge" [33], *cannot* be obtained within the dynamically single-valued, *zero-complexity* projection of the canonical science, *irrespective of details* (similar to any other contradictory *duality*, or "complementarity", of real micro-object behaviour). The same refers to the obtained dynamically unified and causally specified origin of "intrinsic" particle properties, such as mass (inertial and gravitational), energy, electric charge, and spin, which also emerge as different manifestations of the same quantum-beat complexity.

We would like to note a good qualitative agreement of these results, especially the causally extended quantum mechanics, with those obtained by Louis de Broglie in the form of unreduced "double solution" (see [21,23,28-30] for more details and references), and leading "in the same direction" as the dynamic redundance paradigm. Even though de Broglie's theory formally falls within dynamically single-valued analysis and therefore could not provide the truly consistent, dynamically substantiated picture, it is based essentially upon the same irreducibly *realistic* approach in science that characterises the work of Max Planck and actually contains many key details that can be derived in their complete form only within the dynamically multivalued analysis of underlying interaction processes (such as dynamic *nonlinearity*, wave-particle *duality*, and *chaos*, or "hidden thermodynamics of the isolated particle"). By contrast, a number of modern approaches pretending for "causality" and "opposition" with respect to conventional interpretations and often making reference to de Broglie's approach (see e. g. [34,35]), use in reality its degraded versions, which exclude explicit manifestations of unreduced dynamic complexity of the original version and are equivalent to conventional quantum mechanics: they also need to mechanically postulate the existence of observed entities and related "quantum mysteries", without explanation of their true, dynamic origin (see also [22]). Note also that the dynamic multivaluedness phenomenon of quantum field mechanics should be distinguished from any "many-worlds" type of postulated "interpretation" of the standard quantum mechanics, since the dynamically chaotic realisation change occurs precisely due to *uniqueness* of the real world containing the *whole*, redundant set of realisations of any its object.

**(2)** *<u>The number and physical origin of spatial dimensions and time</u>*. Among the obtained results we emphasize the fact that quantum field mechanics provides causal explanation of the *number* (three) and exact *origin* of spatial, physically "tangible" dimensions of actually observed reality and *one* irreversibly flowing, non-material time related to space only indirectly, by system dynamics. Three spatial dimensions originate from three interaction participants (two protofields and their "coupling") whose number is conserved in any their interaction-induced "mixture" (dynamic entanglement), according to the universal complexity conservation law [23], whereas the single time flow marks the unceasing emergence of entanglement *events* explicitly derived as dynamic reduction/extension cycles of the quantum beat process (section 2.2).

Worlds with larger numbers of dimensions and times are possible, in principle, but they should result from more involved starting configurations of underlying interaction and will be qualitatively more complex than our world in their observed behaviour. Such "higher excited states" of the unified world construction can naturally "decay" into simpler, "ground-state" configurations, such as our three-dimensional world with one time (which can well be the most stable or even unique ground state type). Note, however, that this hypothetical splitting of a higher-complexity universe into simpler components is a completely causal, physically real process that has nothing to do with abstract "compactification" of "extra" dimensions artificially inserted into the canonical field theory and absolutely indispensable even in the simplest case of three observable space dimensions (the additional dimensions serve actually as an artificial *substitute* for the *absent dynamic complexity*, similar to many other formal constructions of the canonical theory, including "large", non-compact extra dimensions). Another essential deviation from the conventional lore and in particular its standard relativity postulates is the qualitative difference between space and time, both of them being *physically real* and causally derived in the universal science of complexity: the real, tangible space texture and equally real, but unceasingly flowing, immaterial time cannot be directly "mixed", in any sense, in accord with the totality of existing observations and contrary to the concept of (abstract) "space-time manifolds" of the scholar theory, which should, in addition, be "curved" *as a whole* to account for gravitation.

**(3)** *<u>The number, physical origin, and intrinsic unity of fundamental forces</u>*. Another practically important result is the explanation of exact number of types (four) and physical origin of observed "fundamental interaction forces" (between particles), which are naturally and permanently (dynamically) unified from the beginning, in the obtained holistic picture of quantum field mechanics. Indeed, the observed, real world structure emerges in the system of two interacting, *a priori* structureless protofields, in the form of massive elementary field-particles (quantum beat processes) and their interaction. Since the particle-processes appear as dynamical structures of the same, *physically* unified entities, e/m and gravitational media, they naturally interact with each other by way of direct, "mechanical" tension transmission through each of the media. Interaction of quantum beat processes through the e/m protofield gives rise to the e/m interaction force (transmitted by exchange of real, not "virtual", photonic excitations of e/m protofield), while their similar (somewhat less direct) interaction through the gravitational protofield explains the (naturally quantized!) universal gravitation, in its causal "relativistic" version (including "time retardation effect", see section 2.3). But that kind of "long-range" transmission of tension through the protofields is possible only due to the "short-range" interaction between the neighbouring structural elements of the protofields, which are as real as the protofields, but are at the border of observation from the inside of this world or even slightly beyond it. The short-range interaction between the e/m protofield elements is known as "weak" interaction force, while similar short-range interaction between the gravitational protofield elements is observed as the "strong" force.

It follows immediately that the "elements" of the gravitational protofield appear (only indirectly) in observations as "quarks", which permits us to still decrease the number of unknown entities and points at the direct relation between the "strong" and "gravitational" interaction types. Namely, strong interaction between neighbouring quark elements of the gravitational protofield serves as a particular "microscopic" mechanism for gravitational interaction transmission at longer distances, so that the strongest and the weakest interaction forces are actually unified in the coupled, "gravi-strong" interaction occurring through the gravitational protofield at very short ("strong" force) and larger ("gravitational" force) distances. In a similar fashion, the local ("weak") and nonlocal ("e/m") interaction forces transmitted through the e/m protofield are unified within "electro-weak" interaction, which is only now obtains its *causal* explanation.

The origin of the number (four) and physical essence of observed fundamental interactions becomes thus quite transparent: one should inevitably have two intrinsically unified interaction types for each of the two participating (and interaction-transmitting) media, and $2 \times 2 = 4$ (the result that appears to be inaccessible to the scholar science approach). Other interaction types could in principle exist, so far as different types of, for example, long-range perturbations can exist within protofields, but those four interaction types are really "inevitable", and that is why they should at least be dominating. Any further refinement should be guided by the same direct relation to the real, physical basis of the obtained picture, which is much better than the actually blind, formal (and therefore indefinitely redundant and technically over-complicated) manipulation of the canonical "field theory" that should postulate the existence of a separate, only mathematically defined "field" or abstract "curvature" of "space-time manifold" to account for each experimentally observed force type or particle species.

Contrary to those irreducible separations of the canonical theory, in quantum field mechanics all the four interaction forces are "naturally", dynamically unified in the single quantum beat process within each (hadronic) elementary particle. That kind of intrinsic, physically real and transparent unity of *all* observed elementary entities is quite different from the currently performed (but inevitably failing) attempts of artificial, abstract and mechanistic, "joining" of simplified mathematical imitations within the canonical "unified field theory". It is evident that in each maximum-compression phase of the quantum beat process we have a dense, irregular "ball" expressing the "corpuscular" aspect of the elementary field-particle, where all the described interaction types are "compressed" into "inseparable" mixture that can be considered as a momentary return of the "primordial" system state where the e/m protofield is not yet separated from the gravitational protofield (such unceasing, transient "returns of the past" is the general property of any real, complex-dynamical interaction process). This physically real "unification of interactions" permanently occurs within the heaviest hadrons at the realistic "compression energy" equal to their rest mass ($\sim 100-1000$ GeV) at the frequency of at least $10^{20}$ times per second. The same process, though maybe in a somewhat less complete form, occurs within moderate-mass, stable hadrons, including nucleons. The quantum beat process within leptons unifies rather only weak and e/m interactions, even though the gravi-strong couple is always present "at a distance". One can compare the described natural, physically transparent emergence and dynamic unification of elementary field-particles and their fundamental interactions with their postulated, basically separated, and formal introduction by simplified, redundantly diverse "models" of the canonical theory accompanied by artificial and equally abstract "unification" (that can never succeed).

The mainly qualitative character of the obtained conclusions is their advantage, rather than a shortcoming (as the canonical science book-keepers would judge), since they do provide the *exact*, consistent and unified explanations for the *observed* properties, where many quantitative results are actually involved (see also below) and can be obtained within further theory refinement. We can see now why exactly (i) we have four fundamental interactions two of which are "delocalised", universal, and relatively, though differently, weak (e/m and gravitational forces), while other two are very localised, very (but also differently) strong, and show a limited universality (the "weak" and "strong" forces), (ii) two of the forces (the e/m and "weak" ones) are related in a "couple" (and what it *really* means), while other two (the gravitational and "strong" ones) remain as if separated (but in reality also form a causally interpreted couple of the "gravi-strong" interaction, quite "symmetric" with respect to the first, "electro-weak" one and "impossible" within the canonical approach), (iii) only one force of the four, gravitational interaction, is totally universal and always attractive (as well as quite "peculiar" in its "disobedience" to all canonical imitations of "quantization" and "unification"), (iv) practically all simplified, "exact" symmetries of the conventional field theory are "broken", many of them very strongly (which is actually fixed only in the form of experimental facts by the canonical concept of "broken symmetry"), etc. Such kind of questions cannot even be consistently posed within the canonical formalism and the answers can never be found, even though they do concern the observed, critically important properties of the world foundation. Moreover, the obtained causal picture is further developed and implies many other, practically important consequences for various applications. Thus, the properties of the gravitational medium, emerging through the observed properties of the *causally* interpreted particles and interactions show that its behaviour is quite different from the e/m protofield and is close to that of a dense/incompressible, dissipative fluid, and in that case the canonical search for standard particle-like and wave-like excitations of the "gravitational field" (its conventional, formal "quantization" and search for the "gravity waves") may turn out to be useless in principle, which implies important changes in the whole orientation of research in particle physics, gravity and cosmology.

picture, which is much better than the actually blind, formal (and therefore indefinitely redundant and technically over-complicated) manipulation of the canonical "field theory" that should postulate the existence of a separate, only mathematically defined "field" or abstract "curvature" of "space-time manifold" to account for each experimentally observed force type or particle species.

Contrary to those irreducible separations of the canonical theory, in quantum field mechanics all the four interaction forces are "naturally", dynamically unified in the single quantum beat process within each (hadronic) elementary particle. That kind of intrinsic, physically real and transparent unity of *all* observed elementary entities is quite different from the currently performed (but inevitably failing) attempts of artificial, abstract and mechanistic, "joining" of simplified mathematical imitations within the canonical "unified field theory". It is evident that in each maximum-compression phase of the quantum beat process we have a dense, irregular "ball" expressing the "corpuscular" aspect of the elementary field-particle, where all the described interaction types are "compressed" into "inseparable" mixture that can be considered as a momentary return of the "primordial" system state where the e/m protofield is not yet separated from the gravitational protofield (such unceasing, transient "returns of the past" is the general property of any real, complex-dynamical interaction process). This physically real "unification of interactions" permanently occurs within the heaviest hadrons at the realistic "compression energy" equal to their rest mass ($\sim 100-1000$ GeV) at the frequency of at least $10^{20}$ times per second. The same process, though maybe in a somewhat less complete form, occurs within moderate-mass, stable hadrons, including nucleons. The quantum beat process within leptons unifies rather only weak and e/m interactions, even though the gravi-strong couple is always present "at a distance". One can compare the described natural, physically transparent emergence and dynamic unification of elementary field-particles and their fundamental interactions with their postulated, basically separated, and formal introduction by simplified, redundantly diverse "models" of the canonical theory accompanied by artificial and equally abstract "unification" (that can never succeed).

The mainly qualitative character of the obtained conclusions is their advantage, rather than a shortcoming (as the canonical science book-keepers would judge), since they do provide the *exact*, consistent and unified explanations for the *observed* properties, where many quantitative results are actually involved (see also below) and can be obtained within further theory refinement. We can see now why exactly (i) we have four fundamental interactions two of which are "delocalised", universal, and relatively, though differently, weak (e/m and gravitational forces), while other two are very localised, very (but also differently) strong, and show a limited universality (the "weak" and "strong" forces), (ii) two of the forces (the e/m and "weak" ones) are related in a "couple" (and what it *really* means), while other two (the gravitational and "strong" ones) remain as if separated (but in reality also form a causally interpreted couple of the "gravi-strong" interaction, quite "symmetric" with respect to the first, "electro-weak" one and "impossible" within the canonical approach), (iii) only one force of the four, gravitational interaction, is totally universal and always attractive (as well as quite "peculiar" in its "disobedience" to all canonical imitations of "quantization" and "unification"), (iv) practically all simplified, "exact" symmetries of the conventional field theory are "broken", many of them very strongly (which is actually fixed only in the form of experimental facts by the canonical concept of "broken symmetry"), etc. Such kind of questions cannot even be consistently posed within the canonical formalism and the answers can never be found, even though they do concern the observed, critically important properties of the world foundation. Moreover, the obtained causal picture is further developed and implies many other, practically important consequences for various applications. Thus, the properties of the gravitational medium, emerging through the observed properties of the *causally* interpreted particles and interactions show that its behaviour is quite different from the e/m protofield and is close to that of a dense/incompressible, dissipative fluid, and in that case the canonical search for standard particle-like and wave-like excitations of the "gravitational field" (its conventional, formal "quantization" and search for the "gravity waves") may turn out to be useless in principle, which implies important changes in the whole orientation of research in particle physics, gravity and cosmology.

**(4)** *<u>Causal modification of Planckian units and explanation of observed mass spectrum</u>*. Quantum field mechanics provides [30] a well-substantiated modification (or "renormalisation") of "Planckian units" proposed originally by the author of the hypothesis of quanta [18]. Usual Planckian units obtained by formal combination of universal constants possess a strangely "excessive" values (too small for time and space and too large for mass-energy), which are separated from the corresponding quantities for observed particle species, more than sufficient for the world construction, by an extremely large, apparently "empty" gap (this contradiction is called sometimes "the hierarchy problem"). Thus, the Planckian mass unit is $\sim 10^{17}$ times greater than the largest pseudo-elementary particle mass (including even atomic nuclei). It means that the real world is $10^{17}$ times smaller in its observed ranges of elementary entities (and thus probably in the resulting diversity, apparently quite sufficient), than it should be according to conventional theory estimates: a rather senseless "inflation"!

The interacting protofield picture provides a physically transparent problem solution [23,30]: the combination of the gravitational constant, Planck's constant and light velocity entering the expressions for Planckian units describes in reality the "rough" dynamics of the quantum beat process, and therefore the "gravitational constant" used in *that context* corresponds to the *fundamental attraction between the protofields* and *not* to the *secondary* effects of ordinary gravitational attraction between two field-particles (or many-particle bodies) through the gravitational medium, with its measured, usual value of the gravitational constant. If one replaces the Newtonian gravitational constant by the causally relevant value of the fundamental electro-gravitational coupling, calculated by comparison with the known data and exceeding the conventional value by many orders of magnitude, one obtains just the right values of (renormalised) Planckian units corresponding to the extreme, but *actually observed* properties of elementary particles (thus, the new value of Planckian mass-energy is of the order of $\sim 100$ GeV). This conclusion, based on the dynamically substantiated, realistic picture, leads to considerable modification of various results of the canonical field/particle theory related, directly or indirectly, to usual Planckian unit values. It implies, in particular, a qualitative change of the whole strategy of accelerator research by showing that no "elementary" or "pseudo-elementary" entity can exceed the rest mass of the order of $\sim 100$ GeV. This restriction has a clear interpretation related to the above renormalisation scheme: the maximum possible rest energy simply corresponds to the largest protofield interaction magnitude, as it is expressed by EP amplitude in the respective existence equation (see eq. (2)).

This is only one particular result of the quantum field mechanics, but it clearly demonstrates its general, fundamental difference from the actually blind kind of search of the conventional empiricism: the irreducibly causal, reality-based approach of the universal science of complexity uses mathematics only as a convenient "tool" of expression and computation, but is guided by the *naturally emerging, causally complete understanding* of the *unreduced* (dynamically multivalued) *reality itself*, identical with its *actual development result* from the primal, physically real entities (interacting protofields). In particular, the results and further development of quantum field mechanics provide the objectively "exact", fundamentally substantiated "guiding line" in the "unpredictable" world of particle species and interactions, which shows what and why can or cannot exist in the real world construction (as opposed to practically infinite "liberty" of purely mathematical approach of the canonical field theory giving its current crisis and accumulating "irresolvable" difficulties). The next item provides another example of that "extended causality" of the universal science of complexity.

**(5)** *<u>The origin of mass, inertia-gravity equivalence, and irrelevance of Higgs mechanism</u>*. The causally complete picture of the physical origin of elementary particles and their intrinsic properties, such as mass, shows convincingly that the currently very popular experimental search for the so-called Higgs boson as the "origin of mass" may have very big chances to lead nowhere, simply because any such *artificial* entity inserted by the unitary theory as a "source of mass" is *not necessary* in the realistic world construction, where mass-energy is due to the complex-dynamical (= multivalued, and therefore spatially chaotic) *process* of essentially nonlinear reduction-extension cycles, just constituting the *physical essence* (structure) of real (massive) particles. The *same* process provides, in addition, the dynamic, causal origin of relativity and gravitation [23,28-30] thus *consistently explaining* the postulated "principles" of the canonical theory (including "equivalence between gravitational and inertial masses"), which can hardly be expected from the Higgs mechanism or any other unitary imitation. We would argue that *any* realistic, consistent source of *any* intrinsic, universal enough particle property can only be related to a universal *property* of (inevitably complex) internal *dynamics* necessarily revealing the physical essence of elementary particle, rather than to a special particle species or another artificially added *entity* supposed to directly provide other entities with this property (in the evident contradiction with the *emergent* character of being explicitly observed at all levels of world dynamics). Since the canonical theory cannot propose any version of internal dynamical structure of elementary particles, it is forced to simulate the origin of intrinsic properties by artificial introduction of "responsible" entities, in the form of "special" particles and fields. Therefore, the hypothetical "Higgs" looks as reduced, single-valued imitation of the fundamental protofield interaction "in the whole", which could hardly be detected as such because any real entity is produced by complex-dynamic *development* of this primal interaction, giving the observed variety of particle species and interactions (including also physically real space and time themselves). It could also express causal dynamic unification of the four interaction forces in the "virtual-soliton" phase of the quantum beat process, described above (item 3), but it seems that such interaction unification is not meant in the idea of Higgs boson and can hardly correspond to a particle species. It does not mean that particle species generally "resembling" Higgs boson could not be (often indirectly) "traced" experimentally, but the simplified idea of relating such fundamental (and also macroscopic) property as mass with a special particle/field can lead to misleading interpretations of experimental results. Needless to say, the obtained conclusions also imply important, immediate changes in the performed accelerator research.

**(6)** <u>*Maximum nuclear mass and complex many-body dynamics*</u>**.** The quantum beat dynamics and renormalised Planckian units seem to have interesting relation to the nuclear physics data. Since the nucleus is a complex-dynamic system of *very* strongly interacting and highly *collectivised* particles, it can be considered as a "very big" and somewhat "loose", but still *dynamically unified*, or *elementary* particle. Of course, it is a "compound" one, but such is also any hadron. According to the causally modified meaning of Planckian mass unit, the rest mass of *any* such "pseudo-elementary" species cannot exceed the modified unit value ($\sim 100$ GeV), which explains why atomic nuclei definitely lose their stability at mass values just above the maximum elementary particle mass of $\sim 100$ GeV (note that binding energy contribution and other effects of many-body nuclear dynamics cannot change the order-of-magnitude estimate). One can expect that *any* "compact" (collectivised) ensemble of truly elementary constituents, hold together by short-range forces, cannot exceed in its mass several hundreds of GeV. Let us emphasize that this limitation is based on the most fundamental level of complex world dynamics providing elementary species formation/existence itself, and therefore it does not depend on the details of elementary component interactions at higher sublevels that may provide further refinement of the limiting mass value taking into account particular realisations of that higher-level interaction (giving, for example, more stable or less stable nuclei, etc.).

The applied interpretation of renormalised Planckian energy unit in terms of maximum local protofield interaction energy (item (4)) leads to further fundamental conclusions about the behaviour of quarks and their agglomerates, including causal interpretation of the "confinement" phenomenon. In any case, the behaviour of such *strongly interacting* agglomerates, including atomic nuclei, can be correctly analysed only within the unreduced, dynamically multivalued analysis of the corresponding interaction processes describing a regime of inevitably emerging *genuine* (i.e. *truly random* and purely dynamic) *quantum chaos* [23,26] (see also item 10), as opposed to the canonical, single-valued analysis (including "statistical" description of the basically *regular* quantum "pseudo-chaos", or "chaology" [36,37]).

**(7)** *Causally substantiated strategy of accelerator research and new vision of the universe*. One can mention separately a practically important general side of the above theory applications to particle physics. The obtained unified, causally complete picture of interacting protofields system shows that the coupled protofields represent both the inside and the walls of our largest home called universe. And the local energy "strength" of the walls of our house emerges as a finite value, probably not exceeding too much the value of 100 GeV. This interpretation implies that the standard blind shooting at the wall in the search for an occasional or very "hypothetical", artificially invented entity, performed in the canonical particle physics, is both inefficient for science and potentially dangerous for the world construction. It is inefficient and unreasonable as the number of particle species can well be limited to the current, already redundant diversity of the world structure elements, despite their unlimited diversity in purely abstract, imaginary constructions of conventional field theories (we have shown why they inevitably take a basically wrong way of development). It is potentially dangerous, since the "wall" of the world made essentially of the gravitational medium (represented, most probably, by a "condensed" state of interacting quarks) is also an *intrinsically unstable* complex-dynamical system with a dynamically multivalued, "unpredictable" response to any strong enough perturbation. We do not know either its "thickness" (total "strength"), or detailed reaction to eventual disruption, or the properties of a larger "outside" world, but we know that the strength and any construction "tolerance" are finite, that we have already exceeded its causally substantiated local threshold (contrary to the *wrong* one implied by the conventional Planckian units values, item 4), and that further increase of bombardment intensity can exceed the parameters of naturally occurring perturbations (e. g. by solitary "cosmic rays" or nuclear reactions within stars). We also know that there is a number of super-powerful energy sources in the universe, which remain basically unexplained by usual theory, but can be related to occasional breakdowns of the "normal" structure of the protofield wall of the world.

We deal here also with a qualitatively extended vision of the world structure as compared to the conventional one. The new picture of permanently internally changing, multivalued and probabilistic world construction at *every* its point can be described as "living", and thus "vulnerable", but also full of potentialities, developing and intrinsically *creative* system, as opposed to the canonical *abstract* picture of basically "rigid" matter mechanistically "inserted" into an empty, "fixed" space, even though it may be artificially filled with minor "vacuum fluctuations" (of ambiguous origin). It is evident that usual, effectively one-dimensional and invariably *perturbative* field theory, irrespective of its particular version, *cannot* provide consistent understanding of the *universally nonperturbative*, structure-creating world.

**(8)** *Applicability of classical mechanics to relativistic elementary particle motion*. Essentially quantum effects do not readily appear in both quasi-free motion of relativistic elementary particles in accelerators and their interactions in detectors corresponding rather to the classical type of behaviour with a well-defined trajectory. It is impossible to consistently explain these facts within the conventional theory where "quantum" and "relativistic" properties remain basically separated and even mutually opposed. By contrast, the intrinsically unified description of quantum and relativistic manifestations of quantum beat complexity in quantum field mechanics (section 2.3) provides a natural explanation of that extensively confirmed property: with growing particle energy the proportion of averaged, global-motion component of its dynamics also grows, while the contribution of irregular deviations from that global tendency, just determining "quantum smearing" of the trajectory, respectively decreases [21,23,28-30]. It explains why at higher, "relativistic" energies the "essentially quantum" particle may have a rather narrow, effectively localised trajectory, imitating classical behaviour, even though it preserves its basically quantum, "coherent" properties and can show, for example, undular effects in scattering. In the case of more intense particle interactions in detectors (or any other material "medium"), one should also take into account the unreduced complex dynamics of quantum measurement events [23,27] occurring at each location of strong enough particle-medium interaction. A more detailed experimental study of that "pseudo-classical" behaviour, consistently described by the quantum field mechanics formalism, can provide a direct confirmation of the proposed causal theory of unified quantum and relativistic mechanics and simultaneously give rise to a new, promising direction of accelerator research dealing with detailed analysis of unreduced complex dynamics of elementary particles and their agglomerates, thus avoiding the actually blind and *de facto* fruitless search of imaginary "new species" within the conventional strategy (items 5, 7).

**(9)** *Causal origin of classicality, many-particle diffraction and macroscopic quantum effects*. Independent confirmation of quantum field mechanics comes from its ability to provide a natural, transparent explanation for the origin of "macroscopic" quantum phenomena and undular (diffraction) effects observed for very large molecules and atomic clusters, in contradiction with any conventional, "Hilbert-space" explanation (see ref. [29] for the details). Indeed, because of the formal, non-dynamical postulation of quantum properties, any conventional "explanation" of real, "classical" structure of "ordinary", macroscopic world is forced to postulate simultaneous destruction of "quanticity" due to "decohering" influences of ill-defined, but physically *real* "environment", which are especially dangerous for "vectors" from certain *abstract* spaces. However, even if that "magic" mixture of real and abstract entities could correspond to actually occurring processes, it would imply that "decoherence" should certainly kill any manifestation of unreduced quanticity for larger systems subjected to intense coherence-destroying influences. Observations disproving that simplified picture are too numerous and too diverse to be attributed to particular effects of a quantitative origin. In quantum field mechanics both quantum and classical types of behaviour, as well as the transition between the two, naturally emerge as various regimes and levels of complex, internal *dynamics* of a well-specified *interaction* process between real entities (section 2.3), and therefore it can be shown why for certain cases of that internal interaction the quantum-classical transition can be displaced towards "unusual" persistence of quantum behaviour [29]. Since the quantum-classical transition is but a particular case of "generalised phase transition" between dynamic complexity levels [23] (where classical dynamics emerges as a *higher* complexity level), that irregularly "smeared", but nevertheless well-defined, "quantum-classical border" demonstrates the corresponding general properties of emergence of a new complexity level. Qualitative extension with respect to over-simplified projection of the conventional theory trying to reduce classical behaviour to a (mathematical) "limit" of quantum mechanics, or the reverse, is evident and explains the corresponding "irresolvable" difficulties of the canonical science that become especially frustrating in cases of *explicitly* complex, chaotic and structure-forming, behaviour of many-particle systems (items (10)-(12)).

**(10)** *True quantum chaos, correspondence principle, and the ultimate origin of randomness*. The problem of existence of genuine quantum chaos cannot be solved within the dynamically single-valued approach of scholar quantum mechanics, which is forced to attribute manifestations of chaoticity in Hamiltonian quantum systems to a particularly involved, but fundamentally regular kind of behaviour (see e. g. [36,37]), accompanied by plays of "significant" words, such as "quantum ergodicity", "quantum manifestations of classical chaos" (for the *same*, essentially quantum system), etc. Note that the same situation exists actually in the scholar theory of classical chaos, which attributes randomness in system behaviour to a *regular* mechanism of "dynamical amplification" of random *external* "noise" whose irregularity is actually *postulated* and inserted into the system from the outside. Both this assumption and the mechanism of "exponential trajectory divergence" are incorrect (see ref. [23] for more details), and thus also the attempts of their extension to quantum chaos description. By contrast, the dynamic redundance phenomenon provides the explicit, fundamental source of purely dynamic randomness in *any* real system possessing, however, more evident advantages in the case of Hamiltonian quantum systems, which were clearly demonstrated for real systems of growing universality, from the very beginning of dynamic multivaluedness concept development [23,25-27]. In terms of universal science of complexity, the true quantum chaos belongs to the next higher group of complexity sublevels emerging in interaction processes of elementary particles, which are formed at the lowest complexity sublevels. Another phenomenon from the same sublevel of complexity as quantum chaos is the (causally extended) quantum measurement also resulting from elementary particle interaction, but within a (slightly) "dissipative", open, rather than conservative (Hamiltonian) system dynamics [22,23,27]. Since *any* real interaction gives rise to the true dynamical randomness (that can be more or less pronounced in various systems), the unreduced quantum chaos/measurement description leads to the qualitative, "chaotic" extension [23,26,27] of the whole body of quantum mechanical applications, where the appearing true randomness can often be hidden in empirical "parameters" or purely "statistical", averaged description (cf. the conventional "quantum chaos" theory). As the system complexity grows so that it passes to classical levels of complex dynamics (item 9), the true randomness of quantum dynamics is replaced by the true randomness of classical dynamics, which is actually independent of its quantum counterpart, but has the same fundamental origin (dynamic multivaluedness of the respective interaction processes) and emerges in full agreement with the "correspondence principle" [26], even though the conventional, purely formal version of the latter expresses only simplified, single-valued scheme of the underlying complex-dynamical processes.

**(11)** *Causal origin of radioactive decay and (reversible) chemical reactions*. It is evident that any explicitly "spontaneous", "unpredictable" type of events in many-body quantum systems, such as radioactive decay or reversible chemical reaction, provides direct evidence of dynamical, and *truly random* (multivalued), quantum chaos (items 6, 10). Any regular, however "involved", dynamics could hardly explain the whole observed diversity of phenomena of that kind. The canonical science tends to *postulate* the "probabilistic" character of occurring processes and elementary (often ambiguous) rules for their description. Thus, radioactive decay is "explained" by "quantum tunnelling", which is actually *postulated* itself as another manifestation of "quantum mystery", though endowed with a mathematical representation. The problem exists for tunnelling in *any* quantum system, but for many-particle systems in becomes especially evident that it is rather tunnelling itself that should be *causally described* in terms of system dynamics. Indeed, in quantum field mechanics *every* real interaction potential is an "effective", *dynamically multivalued* one, and visible "quantum penetration" of a particle *as if* "below" the potential barrier is reduced *in reality* to its quite normal passage *above* it, but *only* for those EP realisations that are *lower* than the particle energy in the well. Such realisations *practically always* exist, but appear with various, dynamically determined probabilities (section 2.2), and it is that realisation probability of unreduced interaction process that determines the postulated canonical "tunnelling probability" and probabilistic tunnelling events in phenomena like radioactive decay [23]. Physically, it means simply that there is a probability that interacting system components, moving chaotically, arrange themselves in a particular configuration (or "realisation") where one of them can obtain the impact sufficient e. g. for its separation with the rest of the system, even though that would be impossible in an *average* system configuration (the *only* one considered by the conventional approach). The EP formalism of the universal science of complexity (section 2) simply provides the unified method of derivation of the complete set of system realisations and their probabilities determining, in particular, the probability of tunnelling, radioactive decay, or chemical transformation. Any such process will normally involve a fractal hierarchy of events determining together a (usually) relatively low, but finite probability of the whole process, which can be "classically impossible" only in the dynamically single-valued description of usual theory. Note also that the notion of "chaos-induced tunnelling" often discussed in conventional, always *dynamically single-valued* chaos description is nothing but effectively one-dimensional *imitation* of occurring processes performed in the abstract, "configurational" space (e. g "phase space") and then sometimes silently (and inconsistently) "extended" to real-space phenomena.

**(12)** *Real quantum computers and other micro/nano-devices*. The very popular subject of "quantum computation" (and quantum information processing in general) has a direct connection to the unreduced, causal understanding of the origin of quantum effects, quantum measurement and quantum chaos, since here one should deal with, and even "totally master", the *detailed, internal* dynamics of A.P. Kirilyukquantum processes, contrary to their only "averaged", "statistical" manifestations in many other applications of quantum mechanics, where its standard, inconsistent, but "operationally" efficient "recipes" can be sufficient. Application of unreduced interaction analysis to "quantum computers" gives a situation essentially similar to that of quantum chaos/measurement phenomena [26,27] (item (10)), which means that the canonical, *unitary* quantum computers *cannot* be realised in principle, even in a totally conservative configuration protected from any noise. The reason for that is the same dynamically multivalued realisation emergence in *any* real interaction process killing definitely its unitarity (uniform, single-valued evolution) which is *critically important* just for quantum, nano-scale devices (the assumed unitarity is actually at the very *origin* of expected "magic" properties of quantum computers). Note that efficient "control" of purely dynamic, intrinsic chaos is *impossible* at those lowest complexity levels, since control means interaction, here at the *same* complexity level (contrary to the case of macroscopic devices), that will inevitably produce additional, and relatively *large*, system splitting into redundant realisations leading to *genuine* chaos (which is neglected in the effectively perturbative analysis of such control in the conventional, unitary concept of quantum computers and their stability analysis within the conventional quantum chaos theory). It does not mean that no quantum device can be useful. Any realistic version of quantum machine should be analysed, however, within the unreduced, causally complete description of complex, dynamically redundant interaction processes, which leads to a qualitatively different, dynamically multivalued (and thus "chaotic") type of "computation" [22,23] that cannot be understood within any unitary, abstract-space imitation, or speculative "interpretation", of real quantum dynamics (like "many-worlds" and "quantum histories" interpretations, usual concepts of "quantum entanglement", "quantum teleportation", etc.). Such essentially chaotic dynamics will dominate operation of any real nano-devices, and therefore consistent analysis of these popular systems (known e. g. as "nanomachines") can be performed *only* within the unreduced dynamic redundance paradigm (mostly within its quantum chaos – quantum measurement concept [22,23,26,27]).

**(13)** *Creative universe evolution and cosmology problem solution*. The qualitatively new property of unreduced *creativity* of the universally nonperturbative description within the dynamically multivalued concept of complexity gives rise also to the new, *intrinsically creative* and reality-based *cosmology* considerably modifying the results of canonical, unitary (dynamically single-valued) description of universe evolution and naturally solving its "unsolvable" (and growing) problems. The latter are just related to the basically non-creative, mechanistic nature of the official science concept, equivalent to an effectively one-dimensional (or even zero-dimensional, point-like) projection of the whole diversity of multivalued and multi-level universe dynamics, so that the conventional cosmology is forced to *postulate* the existence/emergence of all its entities and their properties starting from empirical data or intuitive guesses, which does not leave any place for consistent understanding of the true origin of things and their respective emergence. That omnipresent deficiency of canonical science, properly emphasized by Bergson [38], becomes especially evident in the study of explicit *evolution* processes, whether they concern the universe or life on Earth, since the *explicit*, unreduced *creation* of observed complicated structures is absolutely necessary for any evolution to occur. We do not have place here for the detailed description of the fundamental modification brought about by the unreduced dynamic complexity to representation of the universe emergence and development. We may note, however, that quantum field mechanics, as well as the universal science of complexity in general, automatically possesses a naturally "cosmological", "evolutionary" character simply because it is the causally complete theory that describes the *unreduced origin* (nature) of things as they are, which inevitably includes their explicit emergence/derivation from simpler constituents. Thus, we explicitly *obtain* the most elementary entities of the world, its physically real space, time, and elementary particles starting from a primitive structureless configuration of two interacting protofields (see eq. (1)) constituting the irreducible "starting point" of the following quasi-autonomous emergence of universe structure described by the same unified mechanism and formalism of dynamically redundant entanglement of interacting entities [21-23, 26-30]. Growing "irresolvable" difficulties of canonical quantum cosmology, such as the origin of time and mysterious "disappearance" of the "wavefunction of the universe", do not even appear in the unreduced description, since it just shows what is time, how it emerges in the unreduced interaction process [21,23,28-30] and how the initial configuration state-function evolves into a hierarchy of wave- and distribution functions of quantum and classical objects [22] (section 2.3). Since the artificial, mechanically averaged "deformation" of purely abstract "space-time manifold" is replaced, in quantum field mechanics, by the unreduced multivalued dynamics of permanently changing, *living* cosmos, all the related problems of canonical cosmology and gravitation, such as those of "cosmological constant" and "dark matter", also obtain a natural solution.

   ***In conclusion of this section*** it would be not out of place to note that the described qualitatively extended knowledge of the universal science of complexity, oriented to the totally adequate, "exact" representation of reality, shows that one deals here indeed with a new "paradigm" that changes all the "well-established" dogmata of official science proposing only a severely reduced, dynamically single-valued (unitary) projection of reality. In particular, it becomes evident that the conventional way of "theory verification by experiment", actually realised as the search for *subjectively selected*, "point-like" coincidences between *certain* predictions of a totally *abstract* theory and specially arranged, artificial measurements often performed with a technically sophisticated and thus practically ambiguous set-up, *cannot* be efficient in principle for understanding of the irreducibly complex, infinitely diverse reality. Within that canonical approach, one always sees only what one wants to see and can always arrange for any necessary "adjustment" between theory and experiment by playing with complicated systems of parameters and rejecting the undesired "deviations" as unimportant "artefacts". The method of unitary science only *seemed* to work well when it dealt with *specially chosen*, relatively *simple* systems, where explicit manifestations of complexity were relatively weak or rare. However, even those externally "successful" applications are characterised by such glaring inconsistencies as notorious contradictions and supernatural "mysteries" of the "new physics" or total absence of *any* understanding of origin of the most fundamental entities and properties of "classical physics", such as discretely structured space, irreversibly flowing time, mass, energy, gravitation, randomness, etc. The progressively accumulated difficulties culminate in today's profound crisis, or "end" [20], of the *conventional* science. It includes complete devaluation of mechanistic "theory verification by experiment", as it is clearly seen in such fundamental fields as "quantum information processing", "field theory", cosmology, or even solid state physics, where each of the competing, redundantly diverse and purely abstract theories seems to find certain, or even "decisive", "experimental confirmation" before being exposed as containing numerous logical "loops" and other evident inconsistencies behind technically sophisticated abstractions and "special" terminology. Such elaborated concealment of proper inconsistencies (combined with exposing those of other approaches) seems to become the real aim and *actual* result of canonical science at its "ironic" age [20], contrary to the officially announced "search for the truth".

   The above description of *unified* applications of the universal science of complexity (limited here to the lowest complexity levels, cf. [22,23]) demonstrates not only particular results, but qualitatively new general approach to interaction with reality, leading to its *causally complete understanding*. In this "extended causality" approach one looks for the internally complete, totally consistent (non-contradictory) image of reality, even though it is permitted to be somewhat "indistinct". Here one cannot, for example, provide only "separated", mutually incompatible "explanations" for quantum and relativistic effects, or for corpuscular and undular properties of the same object, and call it a "success". The emerging causally complete, or "absolute", reality certainly supposes permanent refinement of details, but at each particular stage of that process one deals with a generally harmonious, coherent, "holistic" image, or *system* with multiple internal connections, which can be summarised simply as unreduced *realism* of the new knowledge, the one that was certainly implied by Max Planck and other *true* founders of "new physics". Correspondingly, "applications" and "experimental confirmations" of the causally complete description of reality cannot be reduced to separate coincidences between results of abstract calculations and subjective experimentation, but should constitute the *totally* consistent picture of all known *natural* phenomena and their aspects, produced by a dense, eventually "non-separable" mixture between "theory" and "experiment". Containing the totality of particular correlations, such approach of the universal science of complexity puts them into a well-defined, properly organised *system* and, concentrating rather on the consistency of the *whole* emerging picture (including all particular details), considers as its final purpose the correct substantiation of *direction* of further *development* of knowledge and production, rather than formal "proof of validity" of any its isolated part as it happens in the unitary science paradigm dealing with infinite series of ever *more* serious contradictions, while the real, *technical* progress advances with the help of *purely empirical*, intuitively guided (and therefore increasingly dangerous) approach. The emerging new kind of *intrinsically unified* knowledge is similar to the holistic process of "image recognition" by the brain, as opposed to a "cutting" kind of cognition within the conventional, dynamically single-valued science.

## 4. Conclusion: The absolute reality returns

Causally complete extension of quantum mechanics naturally unifying it with causally extended relativity and field theory, or "quantum field mechanics", outlined in this paper demonstrates, together with its described applications, the real possibility of solution of the century-old "enigma" of conventional postulates in the direction of totally realistic and truly consistent kind of knowledge always defended by Max Planck, the father of quantum hypothesis. As shown above, all the persisting, "unorthodox" doubts of Max Planck about the conventional, formalistic and obscure, interpretation of quantum phenomena were justified, as well as many of his other "obsessions" like the one around the fundamental nature and role of the second law of thermodynamics or the idea about the necessary unity of scientific world picture. Any progress in the true, causally consistent understanding of reality inevitably approaches us to its unreduced, objectively existing and intrinsically unified version, the Planck's "absolutely reality" that can be specified, as we have seen above, as the unified diversity of the self-developing hierarchy of complex interaction dynamics.

However, history tends to repeat itself and now, at this new century border, the clearly specified, physically and mathematically consistent basis of intrinsically unified, causal understanding of "quantum" (and "classical") reality is demonstratively ignored by the formally dominating adherents of standard abstraction, despite their unceasingly repeated recognition of its basic limitations and the resulting, clearly seen major impasse, or "end", of fundamental physics. The above list of groups of successful applications of quantum field mechanics shows clearly that the true consistency can never be "purely theoretical" and inevitably involves practically important consequences. Since the conventional science development within its dynamically single-valued paradigm, such as "field theory" or vain "quantum" experimentation around the same, unchanged and "mysterious" postulates, provides *no* solution to any real problem at all, this second, "post-modern" version of the same opposition between realism and artificial mystification takes the more and more grotesque forms, as it is demonstrated, in particular, by the series of "quantum" and "millennium" jubilee events and opinions [17,33,39-45] only reproducing once more the same fundamental, evident deficiency of fruitless unitary knowledge.

"One cannot fool all the people all the time", said a wise man, but the history of the "new physics" development seems to be especially designed to provide a counter-example. The true, realistically thinking creators of the new physics, such as Planck, de Broglie, Schrödinger, Lorentz and Poincaré, have revealed the first unambiguous manifestations of unreduced dynamic complexity hidden within the externally "simple" forms of classical world picture and just constituting that clearly "felt" fundamental "novelty" with respect to the old, "classical" physics, but, being unable to provide immediately the consistent, realistic understanding of the emerging new effects, they preferred to continue the search for *that* kind of *causally complete* understanding instead of yielding to the temptation of incomplete, superficial "discoveries". Unfortunately, the destructive spirit of the 20th century gave rise to the massive "new wave" of "revolutionaries" adhering to just the opposite attitude, and the glaring inconsistencies of the "half-made" new physics were simply fixed as "unavoidable" or "practically efficient" postulates and "principles" in exchange to "quick success" (and against the wish of genuine creators of "novelties"), leading to redirection of fundamental science development during the whole century of extremely rapid technical progress into the wrong way of cabalistic, fruitless manipulation with abstract symbols and blind empiricism of trial-and-error experimentation.

However, it is the objective reality, Nature itself that can never be fooled and a civilisation, or "educated" community, trying to prove the reverse will only fool itself and gather bitter fruits corresponding to applied imitations and the time lost in selfish fight of vain ambitions. The subjective effects in science are well known [20,24], and are described, in particular, in Thomas Kuhn's revelation of the structure of scientific revolutions in the modern epoch [46]. All revolutions are not the same, however. Today we have, for the first time in history, the unique situation, where the highly developed, but actually "blind" technology can empirically alter the *whole* scale of natural systems complexity, down to its deepest levels, and *actually* does it, without any consistent understanding of that modified dynamics being instead crudely simplified down to its effectively zero-dimensional, point-like projection (just giving the stagnating "mysteries" and "insoluble" problems). That particular combination of extreme technical power and equally surprising intellectual misery (naively hidden behind self-attributed "distinctions") creates a characteristic dynamic instability of civilisation development also emerging as a unified complex system evolution that cannot be separated any more into simpler parts ("dynamic globalisation" effects), which leads to the critically "sensitive" and important choice between emerging "realisations", or ways, of its further development. Only the unreduced, totally realistic understanding within the intrinsically unified kind of knowledge corresponds to further *complexity growth*, known as *progress*. Various versions of actually dominating, globally enhanced degradation are too evident to be specially emphasised. What remains then is simply to make the choice that should be "personal", but not "subjectively biased", the same kind of choice that was made one hundred years ago by the father of the new physics and was actually always maintained by intrinsic adherents of the unreduced creation.

## References

[1]  M. Planck, "Zur Theorie des Gesetzes der Energieverteilung im Normalspektrum", *Verhandl. Dtsch. phys. Ges.* **2** (1900) 237. See also: M. Planck, "Über das Gesetz der Energieverteilung in Normalspektrum", *Annalen der Physik* **4** (1901) 553.

[2]  M. Planck, *Wissenschaftliche Selbstbiographie* (Leipzig, 1955).

[3]  D. Giulini and N. Straumann, ""..I didn't reflect much on what I was doing.." How Planck discovered his radiation formula", quant-ph/0010008 at http://arXiv.org.

[4]  H. Kragh, "Max Planck: the reluctant revolutionary", *Physics World*, No. 12 (2000) 8.

[5]  M. Planck, "Über eine Verbesserung der Wienschen Spektralgleichung", *Verhandl. Dtsch. phys. Ges.* **2** (1900) 202.

[6]  M. Planck, *Z. angew. Chem.* **42** (1929) 4.

[7]  *The New Physics*, ed. P. Davies (Cambridge University Press, 1989).

[8]  M. Planck, *Phys. Z.* **10** (1909) 62.

[9]  M. Paty, *Eur. J. Phys.* **20** (1999) 373.

[10]  N. Straumann, *Eur. J. Phys.* **20** (1999) 419. Astro-ph/9908342 at http://arXiv.org.

[11]  P.A.M. Dirac, "The Requirements of Fundamental Physical Theory", *Eur. J. Phys.* **5** (1984) 65.

[12]  R.P. Feynman, *The Character of Physical Law* (Cox & Wyman, London, 1965).

[13]  R. Penrose, *Shadows of the Mind* (Oxford University Press, New York, 1994).

[14]  J. Hartle, "Scientific Knowledge from the Perspective of Quantum Cosmology", gr-qc/9601046.

[15]  J. Butterfield and C.J. Isham, "Spacetime and the Philosophical Challenge of Quantum Gravity", gr-qc/9903072 at http://arXiv.org.

[16]  G. 't Hooft, "Quantum Gravity as a Dissipative Deterministic System", *Class. Quant. Grav.* **16** (1999) 3263. Gr-qc/9903084.

[17]  A. Kent, "Night Thoughts of a Quantum Physicist", *Phil. Trans. Roy. Soc. Lond. A* **358** (2000) 75. Physics/9906040 at http://arXiv.org.

[18]  M. Planck, *The theory of heat radiation* (Dover, New York, 1959).

[19]  H. Rechenberg, *Eur. J. Phys.* **20** (1999) 353.

[20]  J. Horgan, *The End of Science. Facing the Limits of Knowledge in the Twilight of the Scientific Age* (Addison-Wesley, Helix, 1996).

[21]  A.P. Kirilyuk, "75 Years of Matter Wave: Louis de Broglie and Renaissance of the Causally Complete Knowledge", quant-ph/9911107 at http://arXiv.org.

[22]  A.P. Kirilyuk, "75 Years of the Wavefunction: Complex-Dynamical Extension of the Original Wave Realism and the Universal Schrödinger Equation", quant-ph/0101129 at http://arXiv.org.

[23]  A.P. Kirilyuk, *Universal Concept of Complexity by the Dynamic Redundance Paradigm: Causal Randomness, Complete Wave Mechanics, and the Ultimate Unification of Knowledge* (Naukova Dumka, Kiev, 1997), 550 p., in English. See also physics/9806002 at http://arXiv.org.

[24]  J. Horgan, "From Complexity to Perplexity", *Scientific American*, June 1995, 74.

[25]  A.P. Kirilyuk, "Theory of charged particle scattering in crystals by the generalised optical potential method", *Nucl. Instr. Meth.* **B69** (1992) 200.

[26]  A.P. Kirilyuk, "Quantum chaos and fundamental multivaluedness of dynamical functions", *Annales de la Fondation L. de Broglie* **21** (1996) 455. Quant-ph/9511034 – 6 at http://arXiv.org.

[27] A.P. Kirilyuk, "Causal Wave Mechanics and the Advent of Complexity. IV. Dynamical origin of quantum indeterminacy and wave reduction", quant-ph/9511037 at http://arXiv.org; "Causal Wave Mechanics and the Advent of Complexity. V. Quantum field mechanics", quant-ph/9511038 at http://arXiv.org; "New concept of dynamic complexity in quantum mechanics and beyond", e-print quant-ph/9805078 at http://arXiv.org.

[28] A.P. Kirilyuk, "Double Solution with Chaos: Dynamic Redundance and Causal Wave-Particle Duality", quant-ph/9902015 at http://arXiv.org.

[29] A.P. Kirilyuk, "Double Solution with Chaos: Completion of de Broglie's Nonlinear Wave Mechanics and its Intrinsic Unification with the Causally Extended Relativity", quant-ph/9902016.

[30] A.P. Kirilyuk, "Universal gravitation as a complex-dynamical process, renormalised Planckian units, and the spectrum of elementary particles", gr-qc/9906077 at http://arXiv.org.

[31] N.F. Mott & H.S.W. Massey, *The Theory of Atomic Collisions* (Clarendon Press, Oxford, 1965).

[32] P.H. Dederichs, "Dynamical Diffraction Theory by Optical Potential Methods", *Solid State Physics: Advances in Research and Applications*, eds. H. Ehrenreich, F. Seitz, and D. Turnbull (Academic Press, New York) **27** (1972) 136.

[33] G. Amelino-Camelia, "Quantum Theory's Last Challenge", *Nature* **408** (2000) 661. Gr-qc/0012049 at http://arXiv.org.

[34] S. Goldstein, "Quantum Theory Without Observers – Part One", *Physics Today*, March 1998, 42. "Quantum Theory Without Observers – Part Two", *Physics Today*, April 1998, 38.

[35] B. Haisch and A. Rueda, *Phys. Lett. A* **268** (2000) 224. Gr-qc/9906084 at http://arXiv.org.

[36] *Chaos and Quantum Physics*, ed. M.-J. Giannoni, A. Voros, and J. Zinn-Justin (North-Holland, Amsterdam, 1991).

[37] B. Chirikov, "Linear and Nonlinear Dynamical Chaos", chao-dyn/9705003 at http://arXiv.org. B. Chirikov, "Pseudochaos in Statistical Physics", chao-dyn/9705004 at http://arXiv.org.

[38] H. Bergson, *L'Évolution Créatrice* (Félix Alcan, Paris, 1907). English translation: *Creative Evolution* (Macmillan, London, 1911).

[39] M. Tegmark and J.A. Wheeler, "100 Years of the Quantum", *Scientific American*, February 2001, 68. Quant-ph/0101077 at http://arXiv.org.

[40] D. Kleppner and R. Jackiw, "One Hundred Years of Quantum Physics", *Science* **289** (2000) 893. Quant-ph/0008092 at http://arXiv.org.

[41] A. Khrennikov, "Foundations of Probability and Physics, Round Table", quant-ph/0101085. See also quant-ph/0302065, quant-ph/0504109, and http://www.msi.vxu.se/aktuellt/konferens/.

[42] "Quantum Theory Centenary", the centenary week and other jubilee events (Berlin, December 2000), http://www.dpg-physik.de/kalender/qt100/qt100e.htm.

[43] "Les Quanta: Un siècle après Planck. Aspects Microscopiques et Macroscopiques de la Physique Quantique", Journée organisée sous l'égide de l'Académie des Sciences (Paris, 15 Décembre 2000), http:// www.sigu7.jussieu.fr/hpr/program.htm.

[44] "75 Years of Quantum Mechanics", jubilee colloquium (Göttingen, 30 October 2000), http://www.theorie.physik.uni-goettingen.de/aktuell/archiv/75qm.en.html.

[45] S. Weinberg, "A Unified Physics by 2050?", *Scientific American*, December 1999, 30.

[46] T. Kuhn, *The Structure of Scientific Revolutions* (Chicago University Press, 1970).